\documentclass[sigconf]{acmart}
\acmSubmissionID{3}

\usepackage{booktabs} 

\usepackage[ruled]{algorithm2e} 

\SetAlFnt{\small}
\SetAlCapFnt{\small}
\SetAlCapNameFnt{\small}
\SetAlCapHSkip{0pt}

\acmJournal{TOG}

\usepackage[inline]{enumitem}
\usepackage{amsmath}
\usepackage{caption}
\usepackage{subcaption}
\usepackage{hyperref}
\usepackage{xcolor}
\usepackage{multirow}
\usepackage{algpseudocode}

\newcommand{\pistrong}[0]{$\pi_{\mathrm{strong}}$~}
\newcommand{\piweak}[0]{$\pi_{\mathrm{weak}}$~}
\newcommand{\pislow}[0]{$\pi_{\mathrm{slow}}$~}
\newcommand{\tauslow}[0]{$\tau_{\mathrm{slow}}$~}
\newcommand{\taufast}[0]{$\tau_{\mathrm{fast}}$~}

\copyrightyear{2022}
\acmYear{2022}
\setcopyright{acmlicensed}\acmConference[SIGGRAPH '22 Conference Proceedings]{Special Interest Group on Computer Graphics and Interactive Techniques Conference Proceedings}{August 7--11, 2022}{Vancouver, BC, Canada}
\acmBooktitle{Special Interest Group on Computer Graphics and Interactive Techniques Conference Proceedings (SIGGRAPH '22 Conference Proceedings), August 7--11, 2022, Vancouver, BC, Canada}
\acmPrice{15.00}
\acmDOI{10.1145/3528233.3530697}
\acmISBN{978-1-4503-9337-9/22/08}

\begin{document}
\title{Learning to Get Up}

\author{Tianxin Tao}
\email{taotianx@cs.ubc.ca}
\affiliation{%
 \institution{University of British Columbia}
 \city{Vancouver}
 \country{Canada}}

\author{Matthew Wilson}
\email{mattwilsonmbw@gmail.com}
\affiliation{%
 \institution{University of British Columbia}
 \city{Vancouver}
 \country{Canada}}

\author{Ruiyu Gou}
\email{reyget42@gmail.com}
\affiliation{%
 \institution{University of British Columbia}
 \city{Vancouver}
 \country{Canada}}

\author{Michiel van de Panne}
\email{van@cs.ubc.ca}
\affiliation{%
 \institution{University of British Columbia}
 \city{Vancouver}
 \country{Canada}}

\begin{abstract}
Getting up from an arbitrary fallen state is a basic human skill.
Existing methods for learning this skill 
often generate highly dynamic and erratic get-up motions, which do not resemble human get-up strategies,
or are based on tracking recorded human get-up motions.
In this paper, we present a staged approach using reinforcement learning, without recourse to motion capture data.
The method first takes advantage of a strong character model, which facilitates the discovery of solution modes. 
A second stage then learns to adapt the control policy to work with 
progressively weaker versions of the character.
Finally, a third stage learns control policies that can reproduce the weaker get-up motions at
much slower speeds.
We show that across multiple runs, the method can discover a diverse variety of get-up strategies, 
and execute them at a variety of speeds.
The results usually produce policies that use a final stand-up strategy that is common to the recovery motions seen from all initial states.  However, we also find policies for which different strategies are seen for prone and supine initial fallen states.
The learned get-up control strategies often have significant static stability, i.e., they can be paused
at a variety of points during the get-up motion.
We further test our method on novel constrained scenarios, such as having a leg and an arm in a cast.

\end{abstract}

%
%

\ccsdesc[500]{Computing methodologies~Animation}
\ccsdesc[500]{Computing methodologies~Reinforcement learning}

%
%

\keywords{Physics-based character animation, learning curriculum}

\begin{teaserfigure}
  \includegraphics[width=0.12\textwidth]{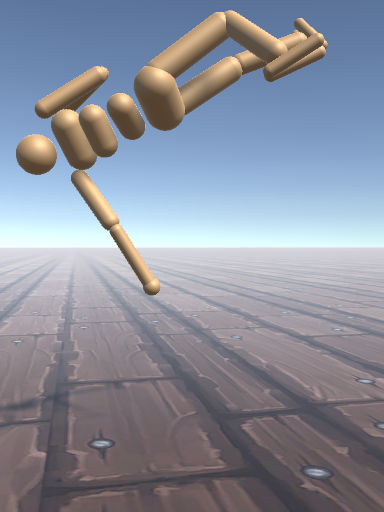}
  \includegraphics[width=0.12\textwidth]{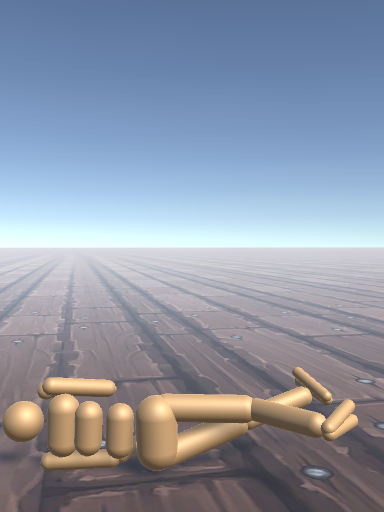}
  \includegraphics[width=0.12\textwidth]{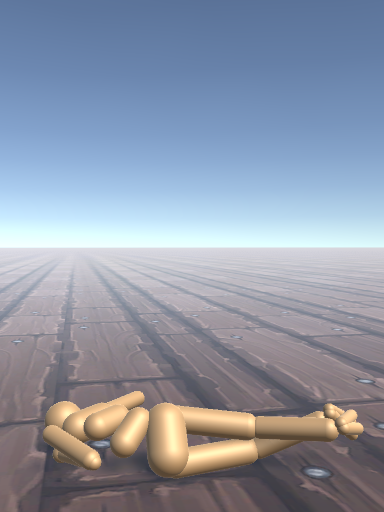}
  \includegraphics[width=0.12\textwidth]{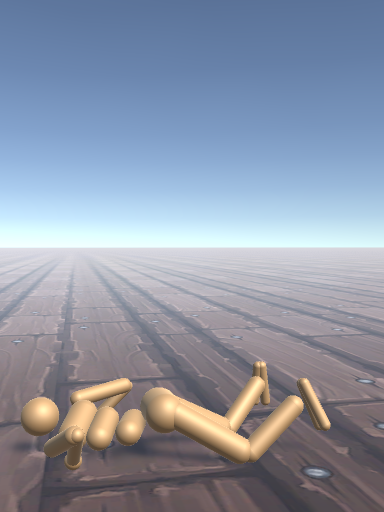}
  \includegraphics[width=0.12\textwidth]{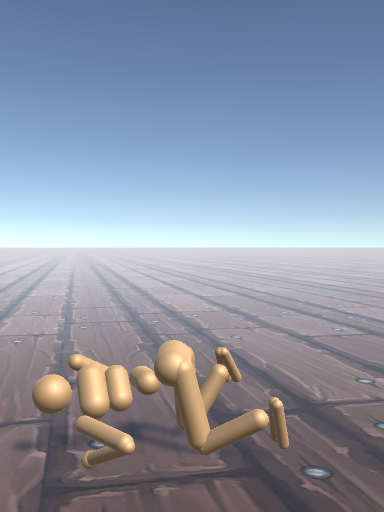}
  \includegraphics[width=0.12\textwidth]{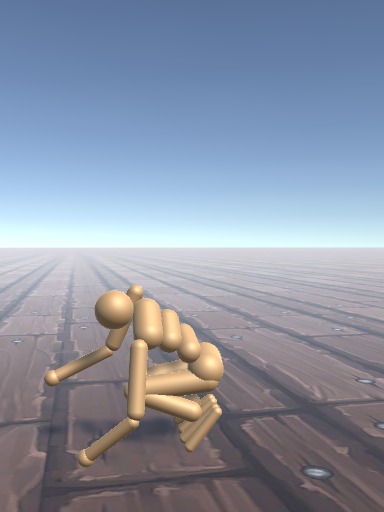}
  \includegraphics[width=0.12\textwidth]{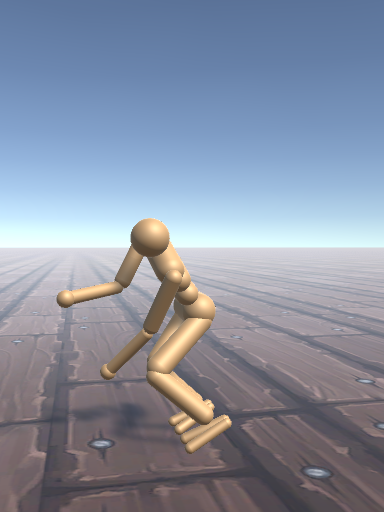}
  \includegraphics[width=0.12\textwidth]{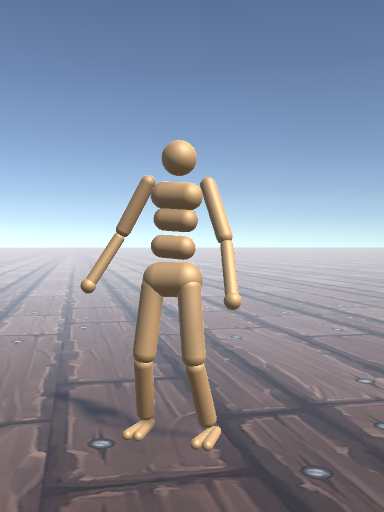}
  \centering
  \caption{We develop a three-stage framework for generating natural and diverse human get-up behaviors using deep reinforcement learning. The learned control policies enable characters to get up from any initial fallen pose, and either slowly or quickly.
  }
  \label{fig:teaser}
\end{teaserfigure}

\maketitle

\section{Introduction}

Getting up from the ground to a standing posture is a natural and effortless skill for most humans.
Simulated characters will similarly need ways of recovering from the arbitrary fallen states
if they are to see broad adoption in simulated worlds.
A common current approach is to learn a policy that simply imitates a relevant motion capture clip,
which thereby bypasses the difficult problem of needing to discover the best get-up strategy.
However, humans get up using a wide variety of styles and speeds, 
and can rapidly improvise when faced with new circumstances, e.g., having a leg in a cast.
This quickly makes it impractical to capture all possible get-up motions in advance.

The get-up problem can also be resolved without recourse to motion capture data, 
and is known as a particularly challenging problem to solve. 
The learned policies, however, often exhibit erratic behaviors
that are unlike those commonly observed in humans~\cite{tassa2020dmcontrol, pinneri2020sample}. 
We speculate that the strong actuation limits made available
in the simulated characters allow for solution modes to be found, while also
leading to these overly dynamic and often-unnatural solutions.

Our work develops a learning framework that achieves more natural get-up motions,
based on the assumption that human get-up motions are typically weak and slow.  
We do this using a three-stage strategy: (i)  learn a get-up control policy for
a strong character, which significantly eases the discovery of solutions modes;
(ii) a curriculum is used to adapt the control policy to a progressively weaker character,
   as implemented via decreasing torque limits;
(iii) a control policy is learned which imitates the outcome of the previous stage at 
speeds up to $5\times$ slower than the original speed.

Our method generates a diverse range of get-up styles across multiple runs.
The discovered motions often have significant static stability, i.e., they can be fully paused at many points in time.  
Learned control policies often exhibit a dominant get-up mode, e.g., 
always first reverting to a prone position before getting up, but can also
exhibit multiple modes, e.g., using different strategies for prone and supine initial states.
We visualize the resulting motion trajectories using t-distributed stochastic neighbor embedding~(t-SNE) plots
to better understand their structure and diversity.  Lastly, 
ablations show the necessity of the various components of our approach. 
For example, we find that directly introducing regularization terms for control effort 
and motion speed leads to a failure to learn.

Our principal contributions are as follows:
\begin{enumerate*}[label=(\arabic*)]
   \item We introduce a method to learn get-up control policies, specifically targeting
   the generation of natural get-up motions without recourse to motion capture data.  
   At the core of our framework is the idea to first discover successful get-up modes,
   and then to learn weak-and-slow versions of these modes.
   \item We visualize and analyze the behavior of multiple learned control policies
   across multiple initial states. This reveals a diversity of strategies, as seen across the controllers arising from multiple runs, as well as for a given controller in response to different initial states.
\end{enumerate*}
We release the code at \url{https://github.com/tianxintao/get_up_control}.

\section{Related Work}


Developing physics-based controllers for character animation is a long-standing research problem. For a thorough history on this topic, we refer the reader to a survey paper~\cite{geijtenbeek2012interactive}. Early work demonstrates significant success on locomotion tasks and often relies on compact manually-designed feedback rules and finite state machines~(FSM), e.g.,~\cite{hodgins1995animating, yin2007simbicon, wang2009optimizing, coros2010generalized, wang2010optimizing}. Trajectory optimization has also been used with considerable success to generate locomotion control for both human and non-human characters~\cite{mordatch2013animating, wampler2014generalizing, kim2021flexible}. Various motion tracking controllers have been proposed to imitate available motion capture data using model-based approaches~\cite{ye2010optimal, lee2014locomotion, lee2010data, yin2007simbicon} or sampling-based methods~\cite{hamalainen2014online,hamalainen2015online,liu2010sampling, liu2015improving, liu2016guided}.

With the rapid progress of deep learning machinery, deep reinforcement learning~(DRL) has become a promising method to learn physics-based controllers. Heess et al.~\cite{heess2017emergence} proposed a framework built on policy-gradient methods to learn a wide range of locomotion skills although the resulting motion suffers from a lack of realism. Many techniques have been proposed to enhance the motion quality. Symmetry constraints have been applied as an inductive bias to produce realistic motions~\cite{yu2018learning, abdolhosseini2019learning}. Yin et al.~\cite{yin2021discovering} proposed the use of pose variational autoencoders in support of learning natural athletic motions. Several works show that designing a curriculum on the task parameters can assist in discovering complicated motions, including dressing and traversing stepping stones~\cite{clegg2020learning, xie2020allsteps}. Alternative actuation models with muscle activation have also been proposed to learn more human-like motions~\cite{lee2019scalable, jiang2019synthesis, peng2017learning}. Additionally, DRL is widely studied to learn tracking controllers for reference motions. Peng et al.~\cite{peng2018deepmimic} proposed a DRL-based framework  with random state initialization~(RSI) and early termination to learn controllers capable of imitating highly diverse motions. DReCon~\cite{bergamin2019drecon} combines the motion matching system and imitation controller to track the motion data.
Building on this, the idea has been extended to address the problem of unlabeled motion data by training a recurrent neural network to predict the reference pose in the next frame~\cite{Park:2019}. 
A hybrid action space of torque and proportional-derivative~(PD) control is proposed to accelerate the training process~\cite{chentanez2018physics}. The methods have also recently been made much more scalable, e.g.,~\cite{won2020scalable}.

Aside from physics-based approaches, kinematics-driven animation methods have also seen significant improvements along with novel deep learning architectures in recent years. Generative animation models are commonly built upon Variational Autoencoders~(VAE)
\cite{ling2020character, rempe2021humor}, Long Short-Term Memory (LSTM) \cite{harvey2018recurrent, harvey2020robust, martinez2017human} and mixture of experts~\cite{zhang2018mode, starke2019neural, starke2020local}. We refer readers to a survey paper for a more detailed overview on this topic~\cite{hoyet2021survey}.

Learning a get-up controller has been of interest to computer animation and robotics. Pioneering work by Morimoto et al.~\cite{morimoto1998reinforcement} proposed a hierarchical reinforcement learning framework to master get-up motions on a simplified 2D walker model. Kanehiro et al.~\cite{kanehiro2003first, kanehiro2007getting} developed a get-up strategy with a manually designed contact graph and careful mechanical calibration to the robot. Bilateral symmetry constraints were proposed to master natural stand-up behavior for humanoid robots~\cite{jeong2016efficient}. A wide range of sampling techniques have been used to successfully discover get-up motions, e.g., ~\cite{hamalainen2014online, pinneri2020sample}. Despite their success, these sampling methods commonly suffer from limited motion quality and are less suited to online use from arbitrary initial states than direct control inference via a DRL control policy. If relevant motion capture data is available, motion tracking methods with DRL are also capable of generating get-up motions, e.g.,~\cite{chentanez2018physics, merel2017learning}. Online trajectory optimization methods, i.e., model predictive control, are also capable of generating get-up motions for humanoids~\cite{tassa2012synthesis}, albeit with limited motion quality and, to the best of our knowledge, restricted to dynamic versions of the motion.
Multiple trajectory optimizations can be structured in a tree-like fashion in order to support reuse 
of the optimization results, for use from a variety of initial states~\cite{borno2017domain}. 
This uses large torque limits (300~Nm) and relies on local PD-control feedback for stability, rather than
closed-loop full state feedback.
In contrast to prior work, we propose a framework which does not need a reference motion, achieves fast runtime performance from arbitrary fallen states on flat ground, and that can produce slow-and-weak motions that are more representative of most human get-up motions.

Human get-up motions are rich and varied in nature, 
with documented demonstrations of at least 52 ways to get up~\cite{52ways},
including a variety of ways to get up without the intermediate use of hands~\cite{no-hands-getup}.
A variety of the methods described require a degree of athleticism and flexibility.  At the other end
of the spectrum is a slow and low-effort strategy described in support of recovery from falls in the elderly,
e.g.,~\cite{adams2000effectiveness}. 
Our work aims to demonstrate how current DRL algorithms can  
learn controllers that can discover get-up strategies that can be at the slower-and-weaker
end of the spectrum of possible strategies.

\nocite{adams2000effectiveness}  
\nocite{no-hands-getup}     
\nocite{52ways}     

\section{Preliminaries}

\begin{figure}
  \centering
  \includegraphics[width=0.95\linewidth]{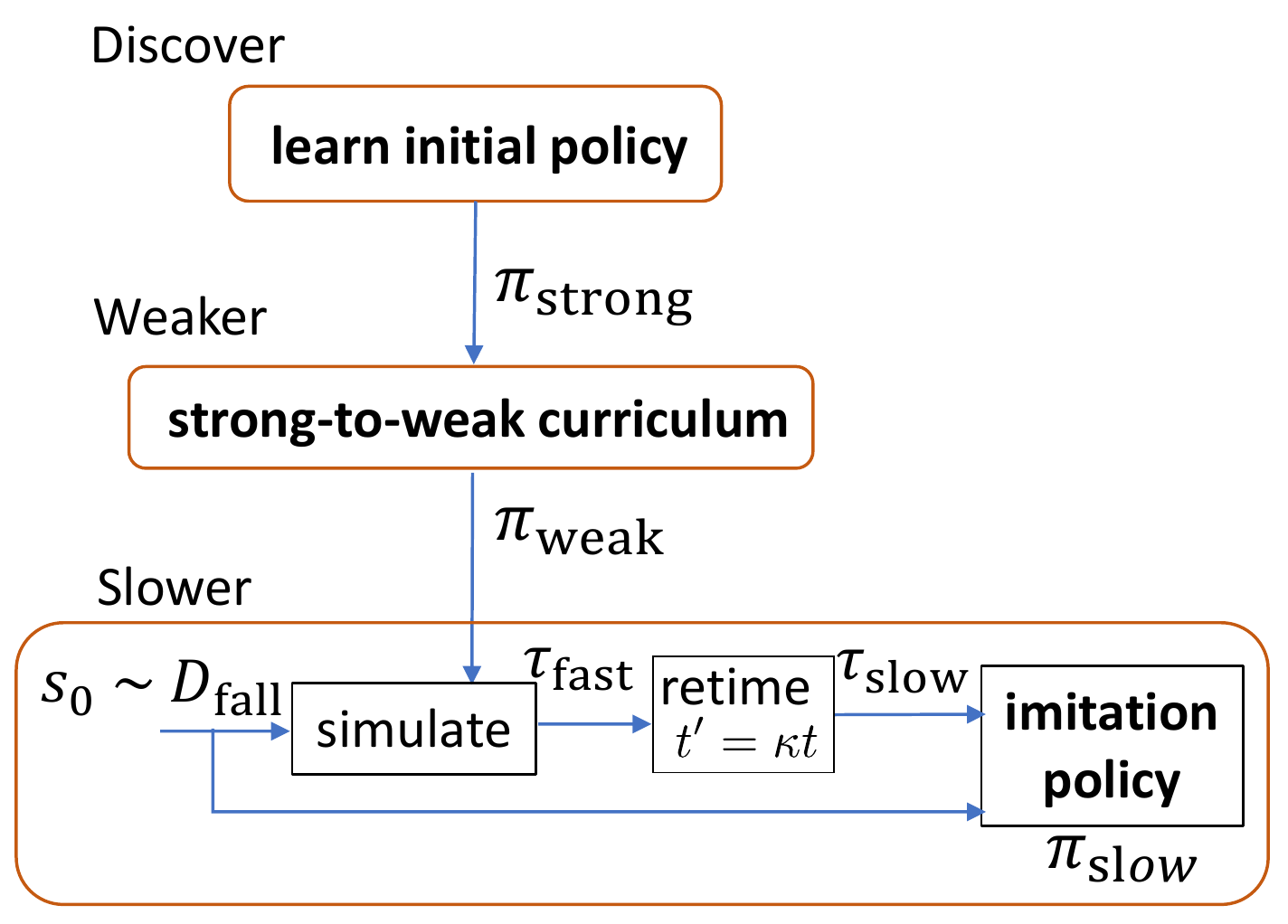}
  \caption{System overview. Our system explores an initial policy with a strong character, then refines the motion from a strong character to a weak character. Finally, we train an imitation policy to track the retimed trajectory produced by the weak policy.}
  \label{fig:overview}
  
\end{figure}

We formulate the DRL problem as a standard Markov Decision Process~(MDP). MDP can be defined by states $s_{t} \in \mathcal{S}$, actions $a_{t} \in \mathcal{A}$, a dynamics function $p(s_{t+1}|s_{t},a_{t})$ denoting the probability of reaching state $s_{t+1}$ with state-action combination $(s_{t},a_{t})$, a discount factor $\gamma \in [0,1]$ and a reward function $R(s_{t}, a_{t})$. The product of DRL is a policy $\pi_{\theta}(s_{t})$ parameterized by $\theta$ interacting with an environment. At each control timestep, the policy selects an action $a_{t}$ given the state $s_{t}$. Then, the agent executes the action $a_{t}$ and the current state $s_{t}$ is transformed into the next state $s_{t+1}$ according to the dynamics function. A scalar feedback $R(s_{t}, a_{t})$ is returned as the reward function. The training objective of DRL is to maximize the expected return as:
\begin{equation}
    J(\theta) = \mathbb{E}_{\tau \sim p_{\theta}(\tau)} \left[ \sum_{t=0}^{T}\gamma^{t}R(s_{t}, a_{t}) \right]
\end{equation}
where $p_{\theta}(\tau)$ represents the probability of experiencing trajectory $\tau = (s_{0},a_{0},s_{1},a_{1},\dots,a_{T-1}, s_{T})$ following policy $\pi_{\theta}$ and T is a finite integer denoting the episode length.

We choose Soft Actor-Critic~(SAC)~\cite{haarnoja2018soft} as the DRL algorithm to train all the tasks in this work. SAC is widely considered as a state of the art of model-free DRL method with excellent sampling efficiency. Besides maximizing the expected return $J_{\theta}$, SAC adopts the idea of entropy regularization to achieve a balance between exploration and exploitation. SAC learns a policy network $\pi_{\theta}(s_{t})$ as actor and an action-value function $Q_{\phi}^{\pi}(s_{t},a_{t})$ as critic, which are commonly represented by multilayer perceptrons~(MLP). In locomotion tasks, the action usually represents the torques of the joint motors, which are commonly normalized to $[-1 ,1]$. To match the action bounds, SAC usually applies a $\tanh$ function on the output as the squashing function.

\section{System Overview}

We illustrate our learning pipeline in Fig.~\ref{fig:overview}. We split the overall training process into three sequential stages: \begin{enumerate*}[label=(\arabic*)]
\item initial policy exploration with strong characters;
\item low-energy motion discovery through a strong-to-weak curriculum; and
\item slow motion refinement via motion imitation.
\end{enumerate*} Discovering get-up motions from scratch using DRL is particularly challenging in that the exploration process can readily become trapped in local minima, which results in variants of a  kneeling motion. To avoid the issue of local minima, we learn an initial get-up controller \pistrong with a character with high torque limits because such a character can explore a larger portion of state and action space. By exploring diverse states and actions, the DRL algorithm is more likely to encounter high reward region to discover a get-up solution mode quickly. However, although a high-strength character is beneficial for exploration, low-strength motions
are usually more natural. To enhance motion quality, we therefore introduce a second stage to progressively learn a policy \piweak suitable for much weaker versions of the character. In practice, we combine the training of the  initial policy \pistrong and its adaption to low-strength actions into one training process. Once the test reward of \pistrong reaches a threshold~$\omega$, the strong-to-weak curriculum is automatically activated.

After the strong-to-weak curriculum, we obtain a state-indexed physics-based controller \piweak generating get-up motions with low energy cost. To generate slower movements, we introduce a motion tracking objective for a third controller \pislow to imitate the retimed trajectories \tauslow. Given an initial state $s_{0}$, we first generate a fast get-up trajectory \taufast using policy \piweak. This is then retimed by a factor of $\kappa, \kappa \in [0,1]$.  The newly trained controller, \pislow, can produce motions up to $5\times$ slower. At run time, users can specify the value of $\kappa$ to adjust the speed of get-up motions. Moreover, we also train the controller \pislow to maintain balance while standing by tracking a manually designed standing pose. 


Our learning pipeline can discover different get-up strategies from prone and supine positions by simply initializing the training with different seeds. We demonstrate and analyze the diverse get-up behaviors using t-SNE plots. We also provide pseudocode for the three stages in the supplementary material.

\section{
`Discover' and `Weaker' Stages}
\label{sec: torque_limit}

In common physics-based simulators, the characters are modelled with torque limits to define the strength of joint motors. Well-designed torque limits play an essential role in the quality of motion. Poorly designed torque limits can lead to unnatural motion~\cite{peng2018deepmimic, merel2018hierarchical} and degraded learning performance~\cite{Abdolhosseini_2019}. Therefore, we design the default torque limits $\mathcal{T}$ according to documented values for humans~\cite{phdthesis}, and scale them with respect to the weights of body parts.

We start the training with the designed torque limits $\mathcal{T}$ of the humanoid character to explore an initial policy \pistrong. Then, the initial policy \pistrong is refined through a strong-to-weak curriculum, where we keep track and update the current torque limits of the character. Once the accumulated minimum test reward of the policy over multiple episodes reaches a specified threshold~$\omega$, we advance the torque limit curriculum by setting the torque limits to be $\beta^{i} \times \mathcal{T}$ at the $i^{th}$ stage of the curriculum, where $\beta\in [0,1]$ is a hyperparameter. To enforce the torque limits on the humanoid character, we design our policy at the $i^{th}$ stage of the curriculum \piweak to output actions bounded by $[-\beta^{i}, \beta^{i}]$ by modifying the squashing function to $\beta^{i} \times \tanh(.)$. Alternatively, the character configuration can be changed along with the curriculum in the simulation, and the action space is always kept to $[-1, 1]$. However, the unchanged action space will confuse the off-policy DRL algorithm because actions collected at different stages of the curriculum then have different meanings.


At the $i^{th}$ stage of the strong-to-weak curriculum, we sample the torque limit multiplier $\beta^{i}$ from a Gaussian distribution $\mathcal{N}(\beta^{i}, \epsilon), \epsilon=0.04$ at the beginning of each episode. The torque limit multiplier is treated as a sampled value rather than a constant because this helps smooth the otherwise discrete nature of the curriculum. 

Previous work commonly employs curriculum on the task objective such as jump heights and stepping stone positions~\cite{yin2021discovering, xie2020allsteps}, which specifies the ultimate task objective with prior knowledge. In our case, the lowest feasible torque limits for diverse get-up strategies are unknown. Thus, we trigger the end of the curriculum based on the number of simulation steps taken at the current stage of the curriculum. Intuitively, the curriculum ends when the current stage requires more gradient steps than a threshold $\mathcal{M}$, which indicates that the current task is too difficult under the torque limit constraints. Rather than a constant, the threshold $\mathcal{M}$ grows as curriculum advances since discovering very low-energy get-up motions becomes more challenging with lower torque limits. We associate the threshold with the number of steps taken at the last stage of the curriculum $N_{i-1}$, and define the threshold at stage $i$, $\mathcal{M}_{i}$ according to $\mathcal{M}_{i} = clip(1.5 \times N_{i-1}, N_{\mathrm{min}}, N_{\mathrm{max}})$, where $N_{\mathrm{min}}$ and $N_{\mathrm{max}}$ are hyperparameters defining minimum and maximum steps for all the curriculum stages.
\section{`Slower' Stage}

To produce slow human-like get-up motions, we introduce a third stage that performs imitation-learning of the linearly retimed version of the trajectories produced by the get-up policy $\pi_{\mathrm{weak}}$.
Specifically, on every episode reset to an initial state $s_{0}$, we iteratively query the get-up policy \piweak to interact with the environment and generate a state trajectory $\tau_{\mathrm{fast}} = (s_{0},s_{1},\dots, s_{T})$ of length $T$. During training, we sample a constant $\kappa$ uniformly between $\kappa_{\mathrm{low}}$ and $\kappa_{\mathrm{high}}$ ($0 < \kappa_{\mathrm{low}} \leq \kappa_{\mathrm{high}} < 1$) as the retiming coefficient for slow get-up trajectories~$\tau_{\mathrm{slow}}$. For retiming, we use linear interpolation on the state trajectory over $[0, T]$. 


To accelerate training, random state initialization (RSI)~\cite{peng2018deepmimic} can be used to initialize an episode from a randomly chosen state on the reference trajectory. We adapt RSI to our acyclic motions using a variant we call $\epsilon$-RSI. $\epsilon$ is a scalar between $[0, 1]$. With probability 1-$\epsilon$, RSI is adopted; otherwise, the episode is started from the beginning. At training time, $\epsilon$-RSI increases the probability of encountering states started from the beginning such that the controller will focus more on achieving the get-up task from end to end. We compare the performance of $\epsilon$-RSI, RSI and without RSI in the supplementary material. In addition, we apply early termination to the episode if the current state diverges too much from the reference motion.

Given a retimed reference trajectory, the policy \pislow aims to imitate it. We choose the PD-controller as the actuation model to compute the joint motor torque. To compute the target orientations $q$, the policy \pislow outputs the a residual value $q_{r}$ added to the reference orientation $q'$ supplied by the reference trajectory~$\tau_{\mathrm{slow}}$: $q=q_{r}+q'$. 
The user can adjust the speed of the get-up motion by controlling the value of the retiming coefficient $\kappa$.

After the get-up motion, the most common and natural succeeding movement is that of quiescent stance. In support of this, we also train the controller \pislow to maintain balance in a natural pose. The training objective is switched to simply track a generic static standing pose after $\frac{T}{\kappa}$ steps into the episode.
\section{Experiment Setup and Task Specification}

We test the proposed method and explore its performance on a regular humanoid model, as well as a modified humanoid model with a leg and an arm in a cast and a humanoid character with a missing arm. 
This involves the design of the rewards for the Discover and Weaker stages, as well as the imitation rewards specific to Slower stage. 
We adopt the reward function design proposed in~\cite{tassa2020dmcontrol} for nearly all the training tasks in this work. 
In the interest of space, we refer the reader to the supplemental material for the low-level details of these generic types of rewards as well as the implementation details.

The get-up task starts from a rag-doll fall at $1.5m$ above the ground with a randomized pose. During the rag-doll fall, actions are randomly sampled according to~$a \sim \mathcal{N}(0, 0.1)$, to model additional stochasticity in the initial states. The rag-doll fall stage lasts for a fixed duration of 80 control steps when the humanoid collides with the ground and remains in a lying pose on the ground afterwards. Then, the controller begins to provide the joint motor torques at each control timestep to accomplish the get-up objective. 

\paragraph{Exploring Initial Policies and its Low-energy Variants.}
The character mainly focuses on finding a coarse get-up solution by maximizing the head height. Each training episode ends when it exceeds $250$ steps without any early termination criteria. The state variable $s_{\mathrm{weak}}$ contains the following attributes:
\begin{enumerate*}[label=(\arabic*)]
\item joint angles and velocities in the local coordinate,
\item head height, center of mass velocity in the world coordinate,
\item end-effector positions in the egocentric coordinate,
\item projection of the torso orientation vector to the z axis of the world coordinate,~$o_{\mathrm{torso}}$,
\item the character strength parameter at $i^{th}$ stage of the curriculum,~$\beta^{i}, 0 \leq \beta \leq 1$.
\end{enumerate*}
The state variable is designed to be agnostic to the facing direction of the humanoid. We further explain the details of the state variables in the supplementary material. The action space is normalized to $[-\beta^{i}, \beta^{i}]$, which is scaled by the default torque limits $\mathcal{T}$ in the simulation to compute the joint motor torques.

The reward function consist of the following terms:
\begin{enumerate*}[label=(\arabic*)]
\item $r_{\mathrm{h}}$: maximize the head height,
\item $r_{\mathrm{straight}}$: keep the torso vertically straight,
\item $r_{\mathrm{v}_{\mathrm{com}}^{\mathrm{xy}}}$: prevent the character from walking or running,
\item $r_{\mathrm{feet}}$: constrain the distance between two feet.
\end{enumerate*}
The detailed explanation for each reward term can be found in App.~\ref{sec:reward_design}. The final reward for the get-up task can be summarized as:
\begin{equation}
    R_{\mathrm{weak}} = r_{\mathrm{h}} \cdot r_{\mathrm{straight}} \cdot r_{\mathrm{v}_{\mathrm{com}}^{\mathrm{xy}}} \cdot r_{\mathrm{feet}}.
\end{equation}

\paragraph{Slow Get-up Motions.} We aim to train a get-up controller receiving different speed commands by specifying the retiming coefficient $\kappa$. We train a motion tracking controller \pislow with DRL by imitating the retimed trajectory~$\tau_{\mathrm{slow}}$. To keep track of the retimed trajectories, we maintain two simulation environments in parallel: one environment that provides the fast reference trajectory for retiming by iteratively querying the previous weak policy~$\pi_{\mathrm{weak}}$, another environment that tracks the retimed slow trajectories~$\tau_{\mathrm{slow}}$. At the beginning of each episode, we obtain a fast get-up trajectory~$\tau_{fast}$ until the head height $h_{\mathrm{head}}$ is above $1.2m$, which is later retimed to a slower trajectory \tauslow through linear interpolation. The training objective is to minimize the distance over a few key attributes between the controlled character motion and the retimed kinematic motion. The maximum length of each episode is determined by the fast trajectory length $T$ and the retiming coefficient $\kappa$ as $\frac{T}{\kappa}$. In addition, the episode will be terminated when the center of mass height deviates from the reference by $0.5m$. 

We design the imitation reward~$R_{\mathrm{slow}}$ consisting of the following elements:
\begin{enumerate*}[label=(\arabic*)]
\item $r_{\mathrm{com}}$: track the center of mass height,
\item $r_{\mathrm{ori}}$: track the torso orientation vectors projected to the vertical axis,
\item $r_{\mathrm{hip}}$: track the hip joint velocity to avoid oscillatory behavior.
\end{enumerate*} The precise definition of each term can be found in App.~\ref{sec:reward_design}.
Thus, the final reward for the imitation phase can be expressed as:
\begin{equation}
    R_{\mathrm{slow}} = r_{\mathrm{com}} \cdot r_{\mathrm{ori}} \cdot \frac{r_{\mathrm{hip}} + 2}{3}
\end{equation}

Furthermore, we train the standing task to maintain balance concurrently with the same policy \pislow by switching to a new reward function $R_{\mathrm{balance}}$ after $\frac{T}{\kappa}$ steps into the episode. The controller is trained to maintain balance for another $100$ timesteps unless early termination is triggered. The early termination will be met if the center of mass height is below $0.5m$. We manually designed a single standing pose~$\hat{q}$ for the imitation policy \pislow to mimic. The designed standing pose~$\hat{q}$ is concatenated to the end of reference trajectory~$\tau_{\mathrm{slow}}$. The reward function for the balancing task mainly reuses reward terms from previous tasks, including $r_{\mathrm{v}_{\mathrm{com}}^{\mathrm{xy}}}$, $r_{\mathrm{straight}}$ and $r_{\mathrm{com}}$. We implement a new reward term~$r_{\mathrm{pose}}$ to track the joint rotations of the designed standing pose. Details of the standing reward design are provided in App.~\ref{sec:reward_design}. The final reward for the rotation task can be expressed as:
\begin{equation}
    R_{\mathrm{balance}} = r_{\mathrm{v}_{\mathrm{com}}^{\mathrm{xy}}} \cdot r_{\mathrm{straight}} \cdot r_{\mathrm{com}} \cdot r_{\mathrm{pose}}
\end{equation}

We also include additional variables in the state space to facilitate the training. In addition to $s_{\mathrm{weak}}$, we select several attributes at two future steps from the reference trajectory \tauslow to augment the state space. The selected attributes are the local joint rotations~$q$, the center of mass height~$h_{\mathrm{com}}$ and the vertical projection of the torso orientation vector~$o_{\mathrm{torso}}$. The concatenation of those attributes forms a vector~$\hat{s}_{t}' = [q_{t}, h_{\mathrm{com_{t}}}, o_{\mathrm{torso_{t}}}]$. The state space at timestep $t$ can be eventually expressed as:~$s_{\mathrm{slow}} = [s_{\mathrm{fast}}, \hat{s}'_{t+1}, \hat{s}'_{t+5}]$. During the get-up phase, the future pose is provided by the retimed slow trajectory \tauslow at the specified timestep, which is later replaced with the standing pose~$\hat{q}$ for the balancing stage. This setup informs the policy of the short-term and long-term goals at the same time.

\paragraph{Get-up Variants.} Besides the standard humanoid character, we also explore and study the generalization ability of our framework on some variants of the humanoid character. We learn get-up motions for a character with a leg and an arm in a cast and a character with a missing arm. To simulate the character with a leg and an arm in a cast, we lock the left elbow joint and the right knee joint to keep those limbs straight throughout the motion. We make no further modifications to the environment and algorithm design except for removing corresponding joint information from the state space and control signal from the action space.


\section{Results}

\begin{figure}
  \centering
   \begin{subfigure}{0.45\linewidth}
    \includegraphics[width=\linewidth]{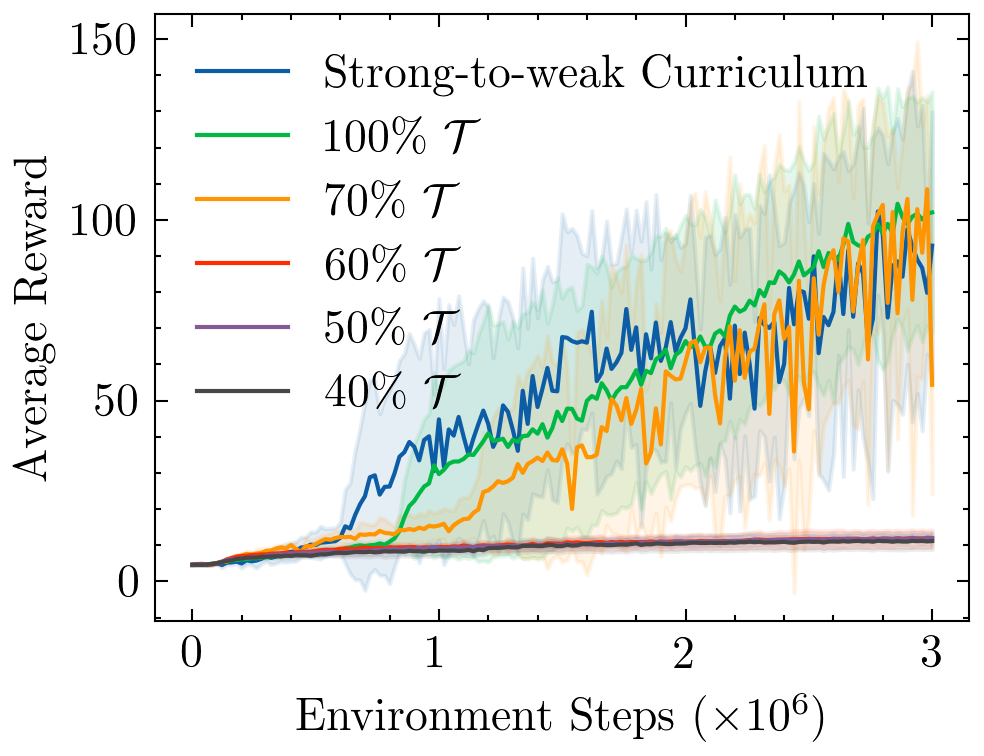}
    \caption{Test reward during training}
    \label{fig:learning_curve}
  \end{subfigure}
  \begin{subfigure}{0.45\linewidth}
    \includegraphics[width=\linewidth]{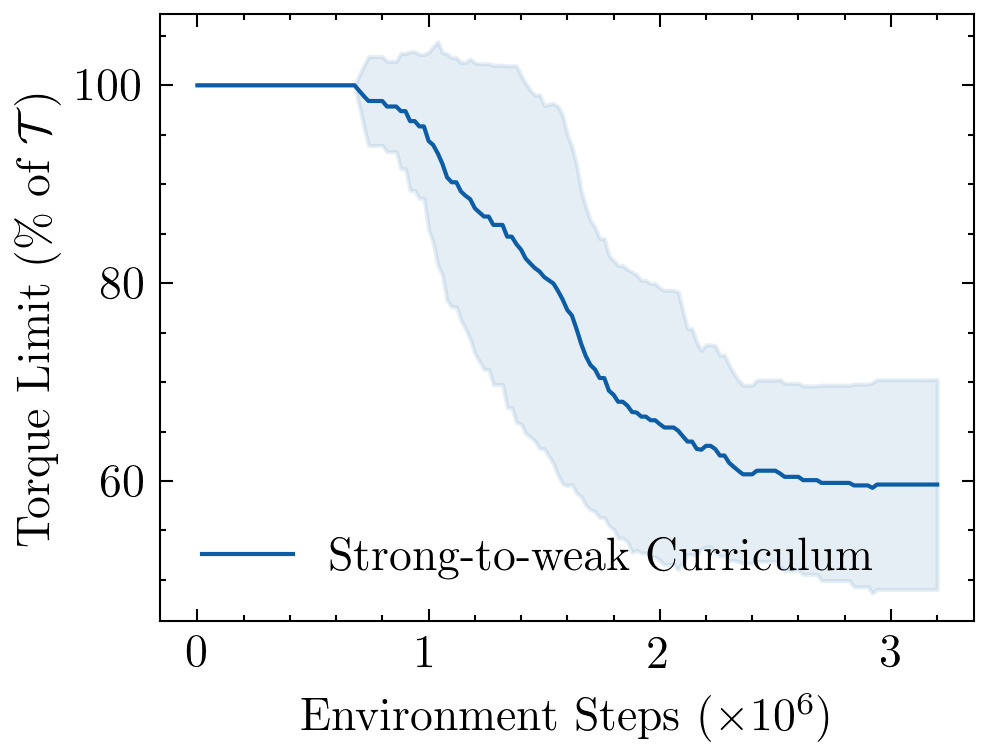}
    \caption{Torque limits during training}
    \label{fig:torque_limit_curve}
  \end{subfigure}
   \caption{a) Average test reward curve for the strong-to-weak curriculum and different values of the fixed torque limits. b) Value of the torque limits in training with a curriculum. The results are averaged over 10 runs.}
   \label{fig:torque_curriculum}
\end{figure}

We first verify the hypothesis that large torque limits are essential to exploration. Low torque limits can prevent the discovery of get-up solutions due to limited exploration. We demonstrate that gradually reducing the torque limits with a curriculum can learn a low-energy solution mode. Then, we show that learning slow get-up motions can further improve the naturalness of the motion. Also, we exploit the future pose conditioned policy \pislow to pause the get-up motions in selected statically-stable poses. Finally, we provide visualization tools to analyse the behavior of each controller and the diversity of the learnt solution modes. We refer readers to the supplementary videos for a clear demonstration of the resulting motions.

\subsection{Strong-to-weak Curriculum}
\label{sec:torque_limit_results}

\begin{figure}
  \def\figwidth{0.115\linewidth}
  \centering
  
  \begin{subfigure}{\linewidth}
      \centering
      \begin{subfigure}{\figwidth}
        \includegraphics[width=\linewidth]{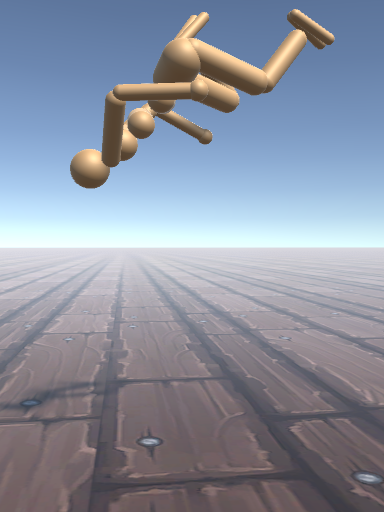}
      \end{subfigure}
      \begin{subfigure}{\figwidth}
        \includegraphics[width=\linewidth]{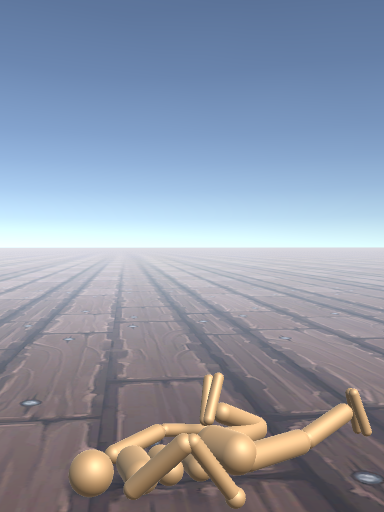}
      \end{subfigure}
      \begin{subfigure}{\figwidth}
        \includegraphics[width=\linewidth]{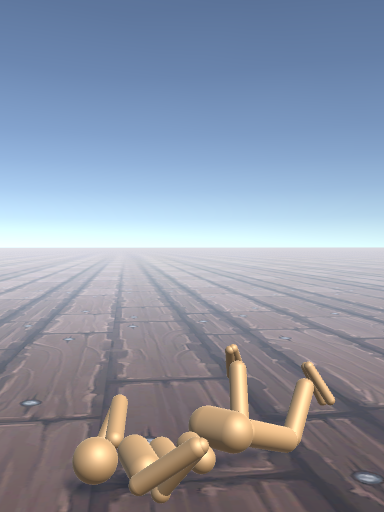}
      \end{subfigure}
      \begin{subfigure}{\figwidth}
        \includegraphics[width=\linewidth]{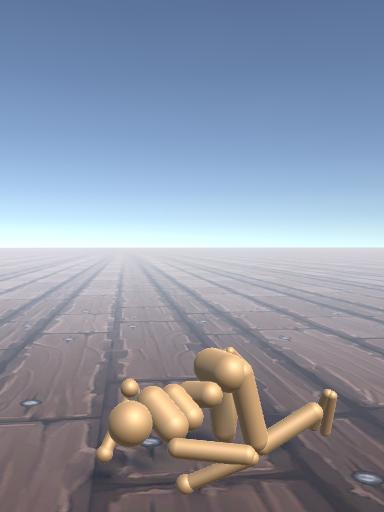}
      \end{subfigure}
      \begin{subfigure}{\figwidth}
        \includegraphics[width=\linewidth]{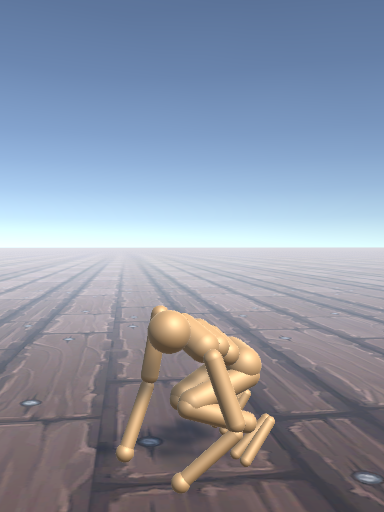}
      \end{subfigure}
      \begin{subfigure}{\figwidth}
        \includegraphics[width=\linewidth]{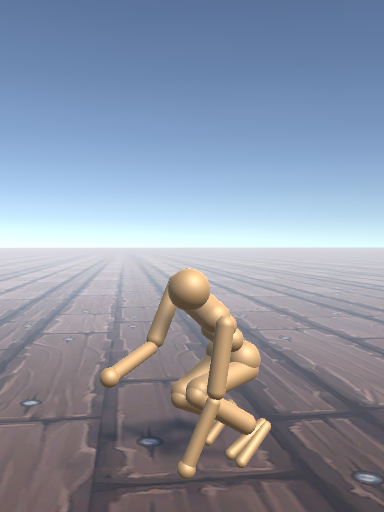}
      \end{subfigure}
      \begin{subfigure}{\figwidth}
        \includegraphics[width=\linewidth]{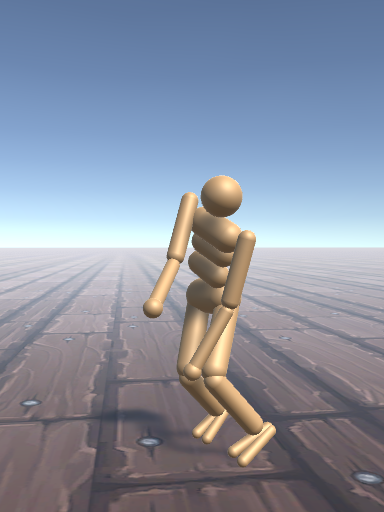}
      \end{subfigure}
      \begin{subfigure}{\figwidth}
        \includegraphics[width=\linewidth]{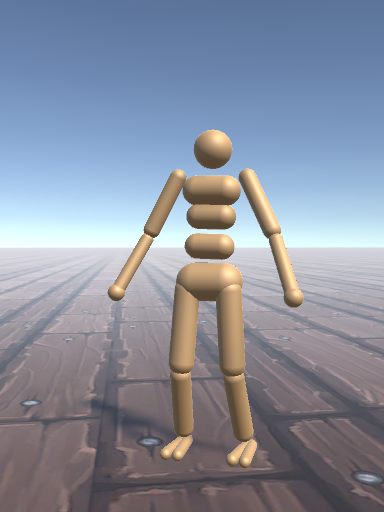}
      \end{subfigure}
      
      \begin{subfigure}{\figwidth}
        \includegraphics[width=\linewidth]{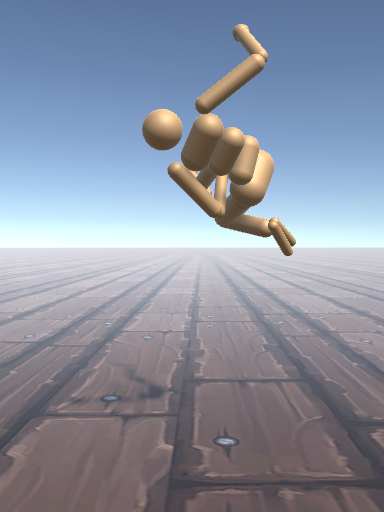}
      \end{subfigure}
      \begin{subfigure}{\figwidth}
        \includegraphics[width=\linewidth]{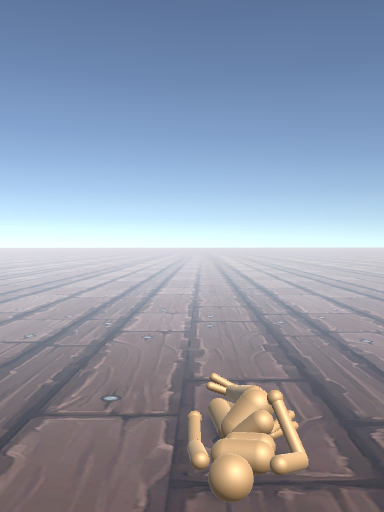}
      \end{subfigure}
      \begin{subfigure}{\figwidth}
        \includegraphics[width=\linewidth]{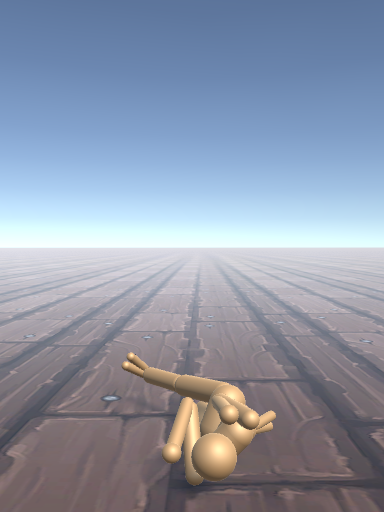}
      \end{subfigure}
      \begin{subfigure}{\figwidth}
        \includegraphics[width=\linewidth]{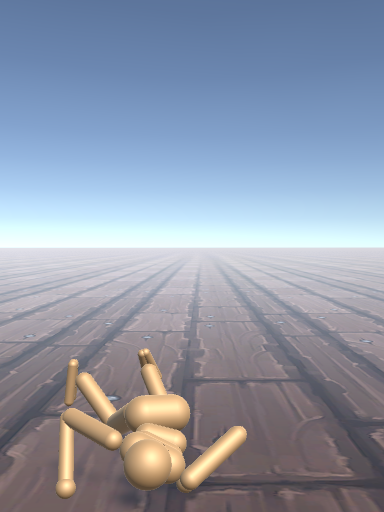}
      \end{subfigure}
      \begin{subfigure}{\figwidth}
        \includegraphics[width=\linewidth]{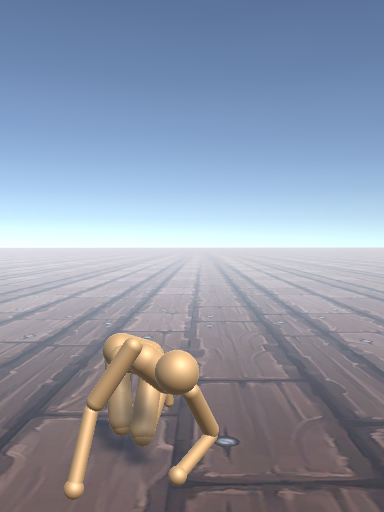}
      \end{subfigure}
      \begin{subfigure}{\figwidth}
        \includegraphics[width=\linewidth]{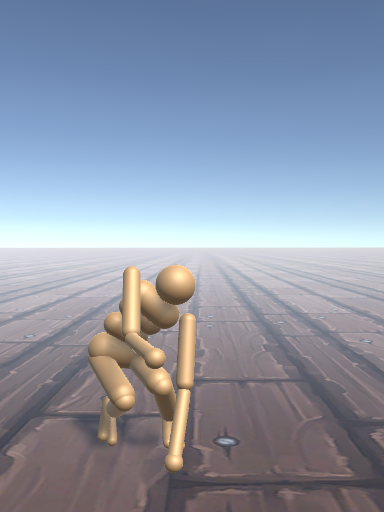}
      \end{subfigure}
      \begin{subfigure}{\figwidth}
        \includegraphics[width=\linewidth]{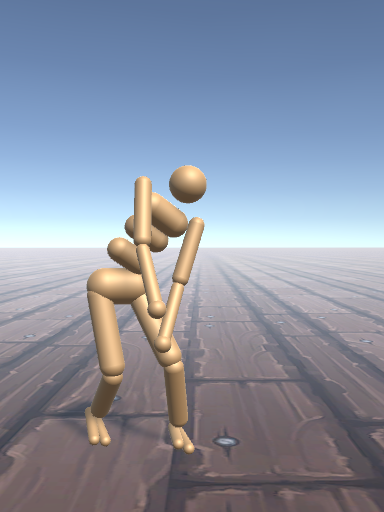}
      \end{subfigure}
      \begin{subfigure}{\figwidth}
        \includegraphics[width=\linewidth]{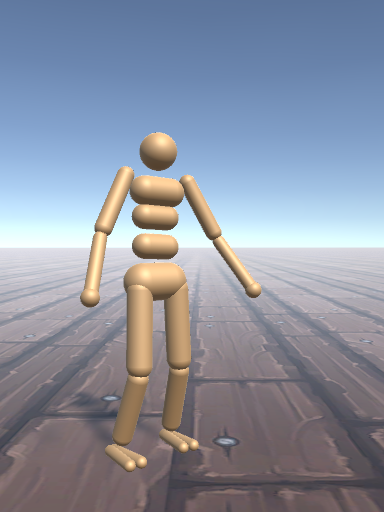}
      \end{subfigure}
      
      \caption{Controller A}
  \end{subfigure}
  
  \begin{subfigure}{\linewidth}
      \centering
      \begin{subfigure}{\figwidth}
        \includegraphics[width=\linewidth]{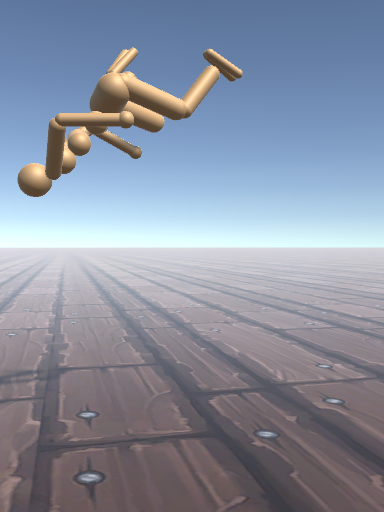}
      \end{subfigure}
      \begin{subfigure}{\figwidth}
        \includegraphics[width=\linewidth]{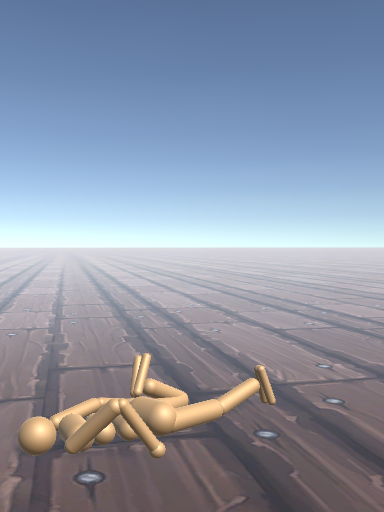}
      \end{subfigure}
      \begin{subfigure}{\figwidth}
        \includegraphics[width=\linewidth]{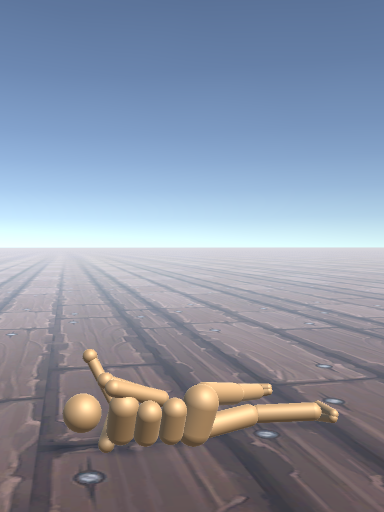}
      \end{subfigure}
      \begin{subfigure}{\figwidth}
        \includegraphics[width=\linewidth]{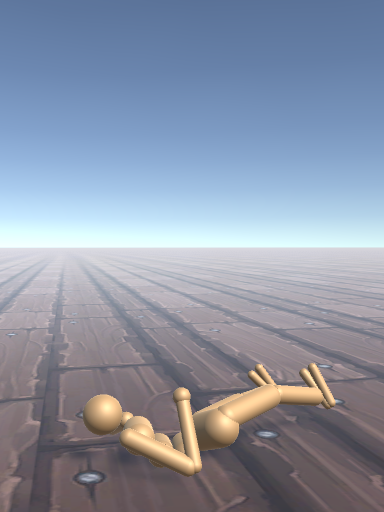}
      \end{subfigure}
      \begin{subfigure}{\figwidth}
        \includegraphics[width=\linewidth]{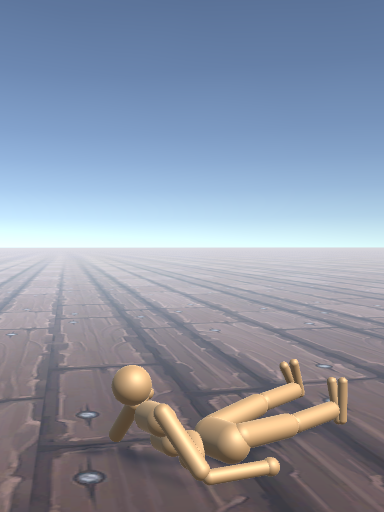}
      \end{subfigure}
      \begin{subfigure}{\figwidth}
        \includegraphics[width=\linewidth]{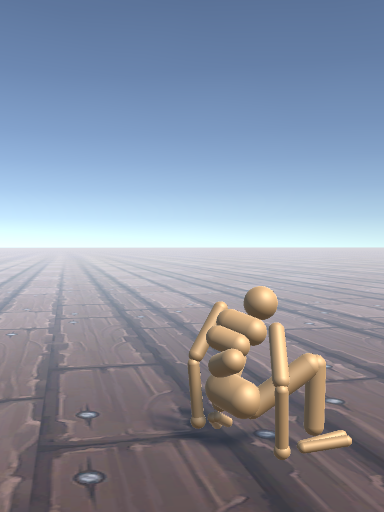}
      \end{subfigure}
      \begin{subfigure}{\figwidth}
        \includegraphics[width=\linewidth]{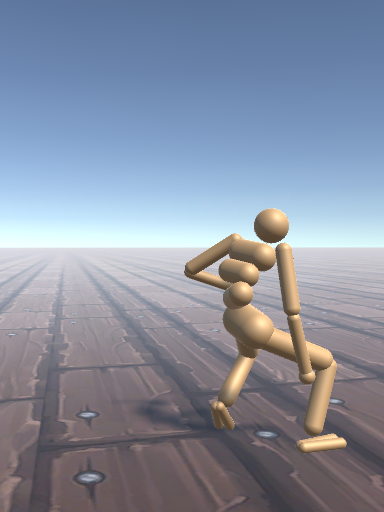}
      \end{subfigure}
      \begin{subfigure}{\figwidth}
        \includegraphics[width=\linewidth]{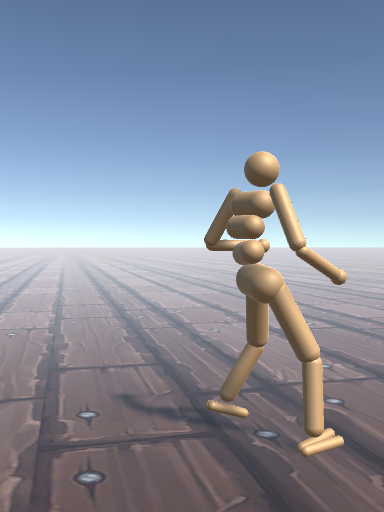}
      \end{subfigure}

      \begin{subfigure}{\figwidth}
        \includegraphics[width=\linewidth]{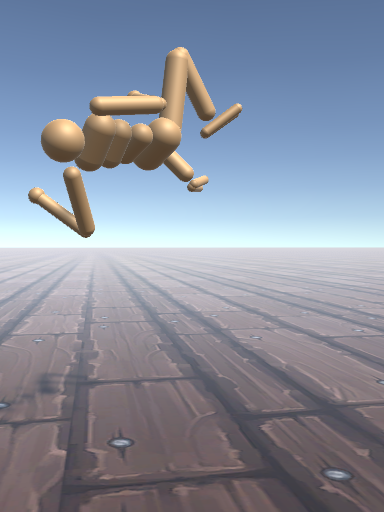}
      \end{subfigure}
      \begin{subfigure}{\figwidth}
        \includegraphics[width=\linewidth]{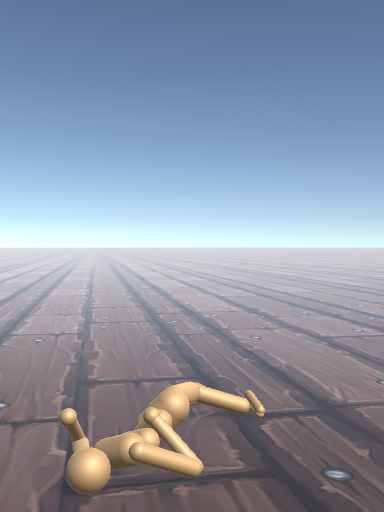}
      \end{subfigure}
      \begin{subfigure}{\figwidth}
        \includegraphics[width=\linewidth]{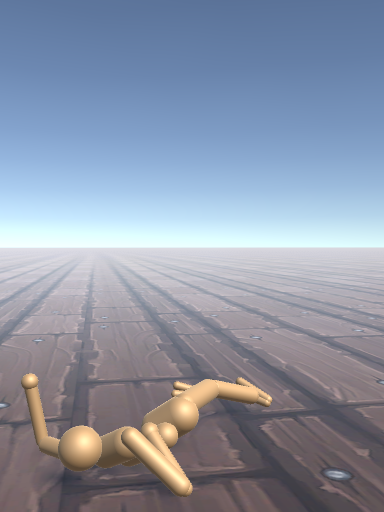}
      \end{subfigure}
      \begin{subfigure}{\figwidth}
        \includegraphics[width=\linewidth]{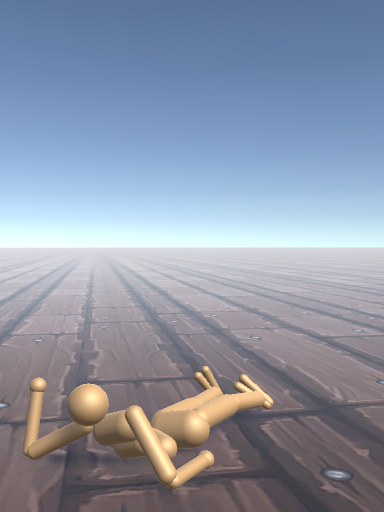}
      \end{subfigure}
      \begin{subfigure}{\figwidth}
        \includegraphics[width=\linewidth]{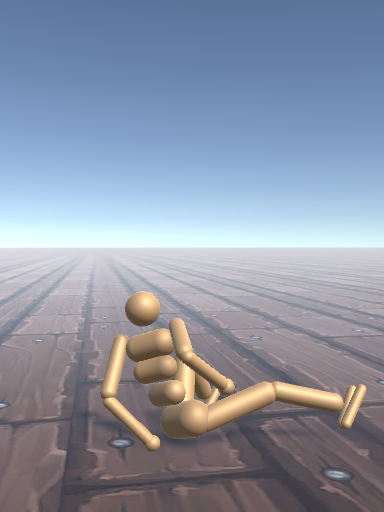}
      \end{subfigure}
      \begin{subfigure}{\figwidth}
        \includegraphics[width=\linewidth]{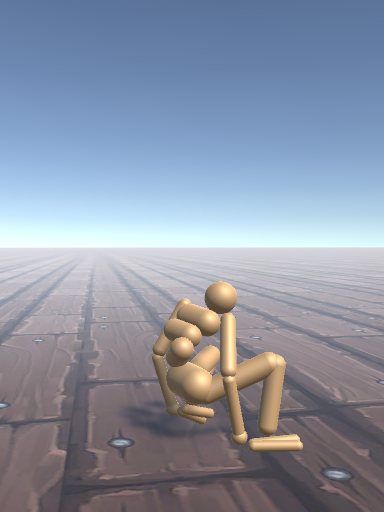}
      \end{subfigure}
      \begin{subfigure}{\figwidth}
        \includegraphics[width=\linewidth]{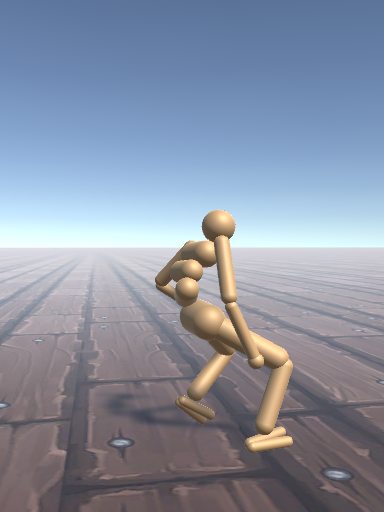}
      \end{subfigure}
      \begin{subfigure}{\figwidth}
        \includegraphics[width=\linewidth]{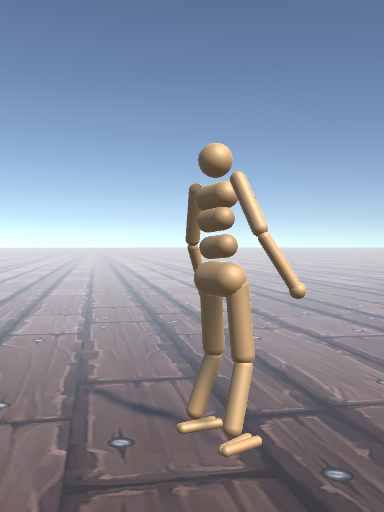}
      \end{subfigure}
      
      \caption{Controller B}
  \end{subfigure}
  
  \begin{subfigure}{\linewidth}
      \centering
      \begin{subfigure}{\figwidth}
        \includegraphics[width=\linewidth]{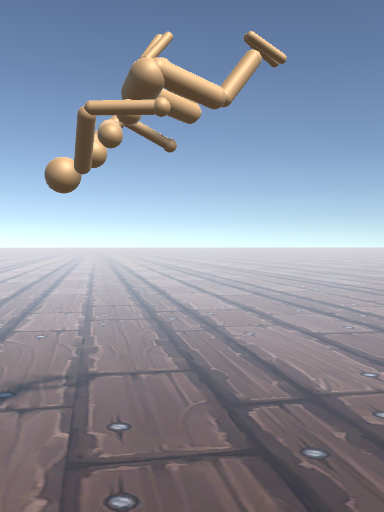}
      \end{subfigure}
      \begin{subfigure}{\figwidth}
        \includegraphics[width=\linewidth]{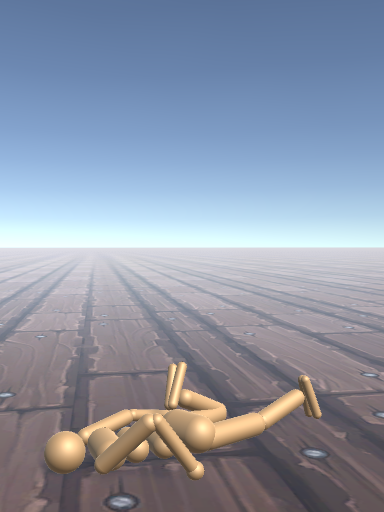}
      \end{subfigure}
      \begin{subfigure}{\figwidth}
        \includegraphics[width=\linewidth]{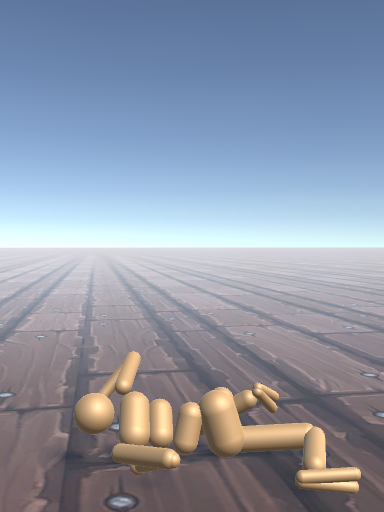}
      \end{subfigure}
      \begin{subfigure}{\figwidth}
        \includegraphics[width=\linewidth]{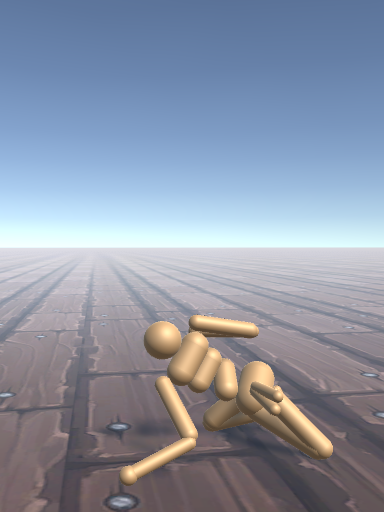}
      \end{subfigure}
      \begin{subfigure}{\figwidth}
        \includegraphics[width=\linewidth]{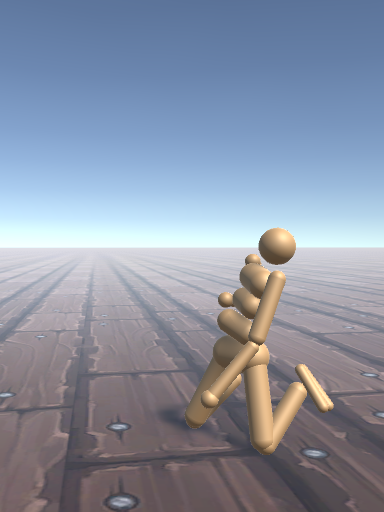}
      \end{subfigure}
      \begin{subfigure}{\figwidth}
        \includegraphics[width=\linewidth]{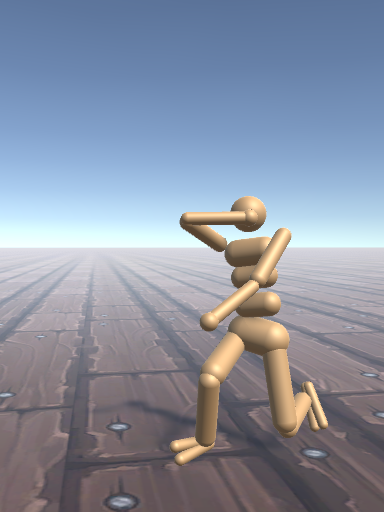}
      \end{subfigure}
      \begin{subfigure}{\figwidth}
        \includegraphics[width=\linewidth]{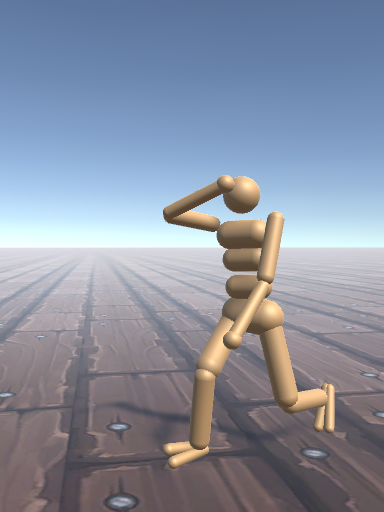}
      \end{subfigure}
      \begin{subfigure}{\figwidth}
        \includegraphics[width=\linewidth]{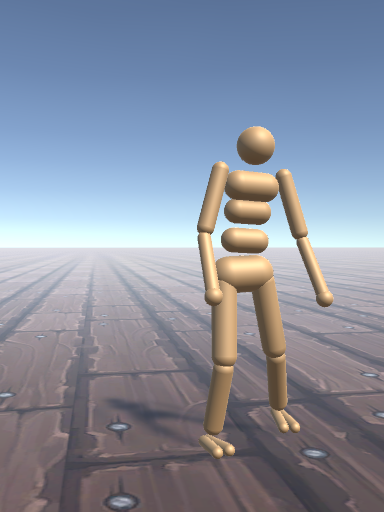}
      \end{subfigure}

      \begin{subfigure}{\figwidth}
        \includegraphics[width=\linewidth]{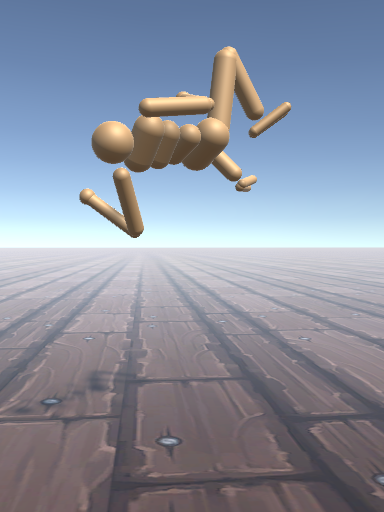}
      \end{subfigure}
      \begin{subfigure}{\figwidth}
        \includegraphics[width=\linewidth]{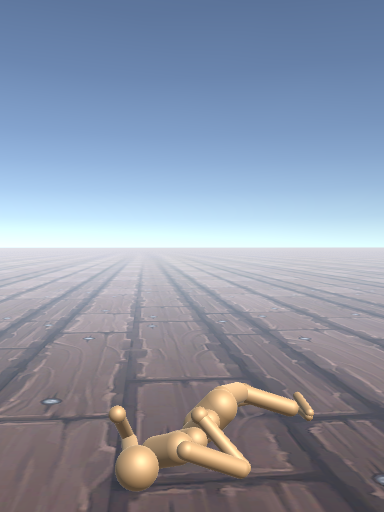}
      \end{subfigure}
      \begin{subfigure}{\figwidth}
        \includegraphics[width=\linewidth]{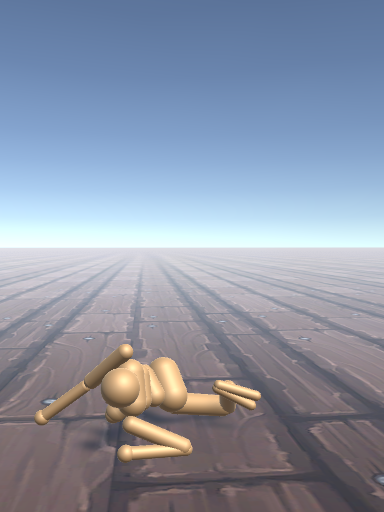}
      \end{subfigure}
      \begin{subfigure}{\figwidth}
        \includegraphics[width=\linewidth]{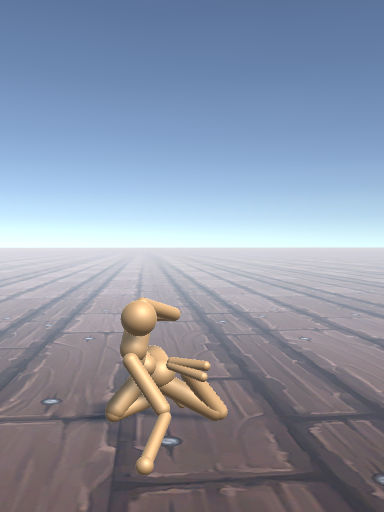}
      \end{subfigure}
      \begin{subfigure}{\figwidth}
        \includegraphics[width=\linewidth]{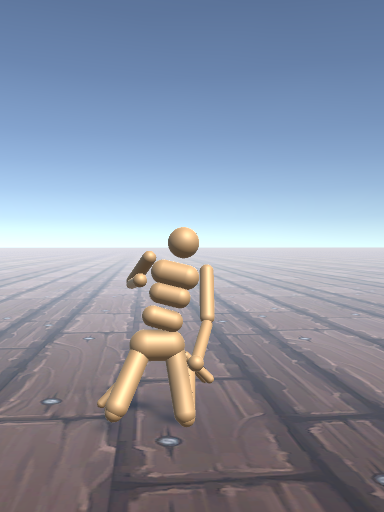}
      \end{subfigure}
      \begin{subfigure}{\figwidth}
        \includegraphics[width=\linewidth]{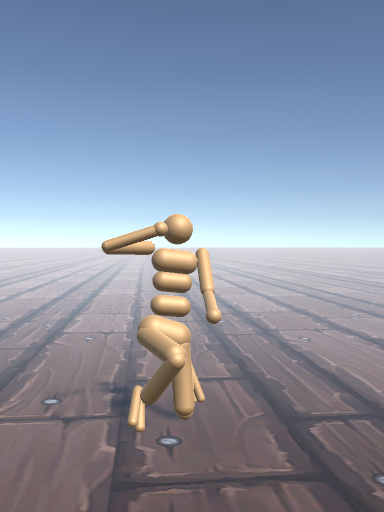}
      \end{subfigure}
      \begin{subfigure}{\figwidth}
        \includegraphics[width=\linewidth]{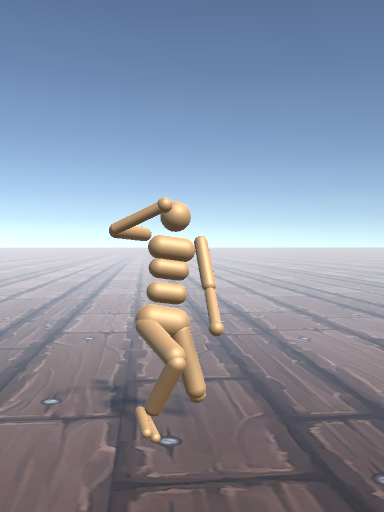}
      \end{subfigure}
      \begin{subfigure}{\figwidth}
        \includegraphics[width=\linewidth]{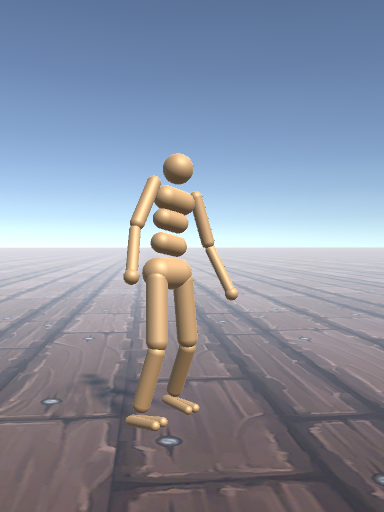}
      \end{subfigure}
      
      \caption{Controller C}
  \end{subfigure}
  
  \begin{subfigure}{\linewidth}
      \centering
      \begin{subfigure}{\figwidth}
        \includegraphics[width=\linewidth]{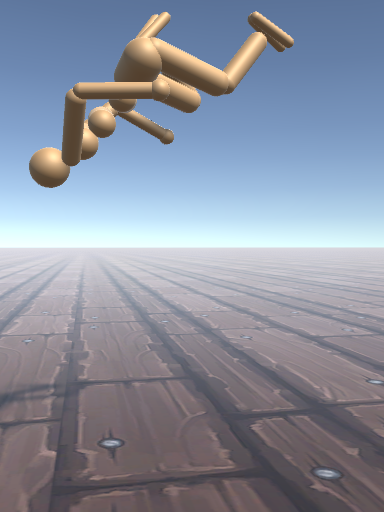}
      \end{subfigure}
      \begin{subfigure}{\figwidth}
        \includegraphics[width=\linewidth]{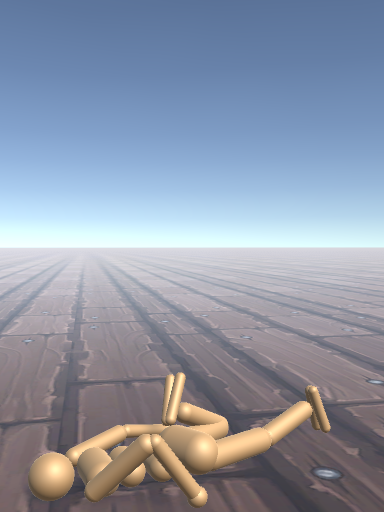}
      \end{subfigure}
      \begin{subfigure}{\figwidth}
        \includegraphics[width=\linewidth]{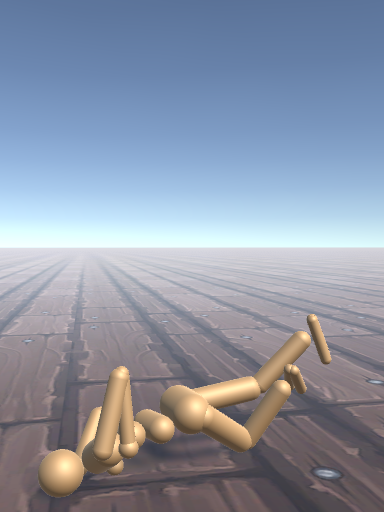}
      \end{subfigure}
      \begin{subfigure}{\figwidth}
        \includegraphics[width=\linewidth]{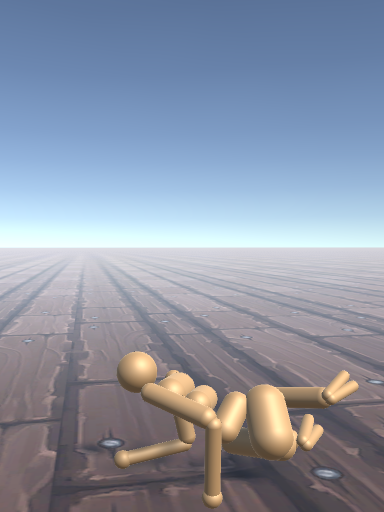}
      \end{subfigure}
      \begin{subfigure}{\figwidth}
        \includegraphics[width=\linewidth]{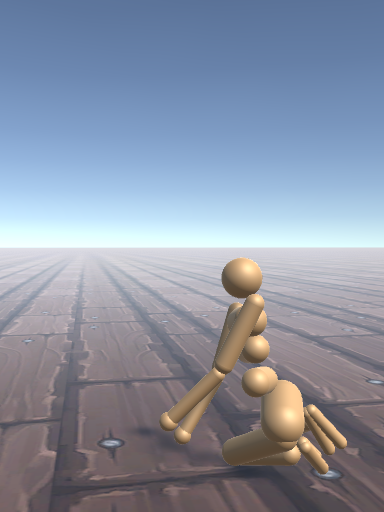}
      \end{subfigure}
      \begin{subfigure}{\figwidth}
        \includegraphics[width=\linewidth]{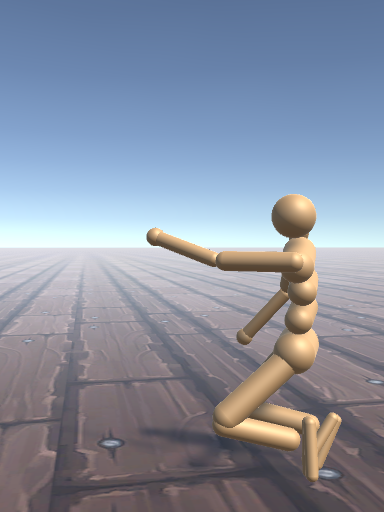}
      \end{subfigure}
      \begin{subfigure}{\figwidth}
        \includegraphics[width=\linewidth]{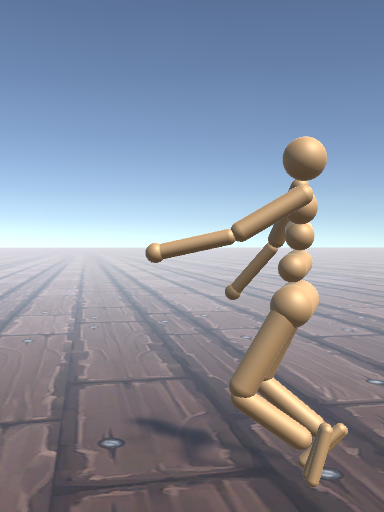}
      \end{subfigure}
      \begin{subfigure}{\figwidth}
        \includegraphics[width=\linewidth]{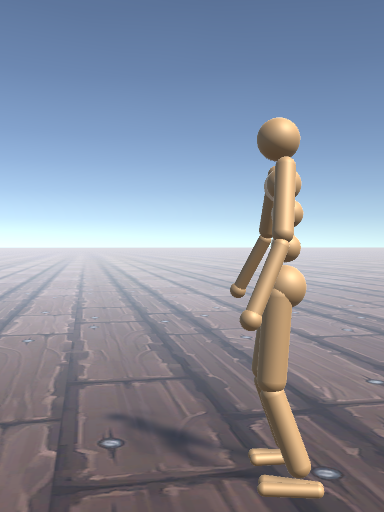}
      \end{subfigure}

      \begin{subfigure}{\figwidth}
        \includegraphics[width=\linewidth]{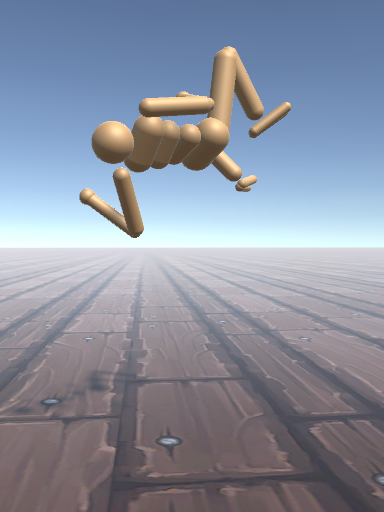}
      \end{subfigure}
      \begin{subfigure}{\figwidth}
        \includegraphics[width=\linewidth]{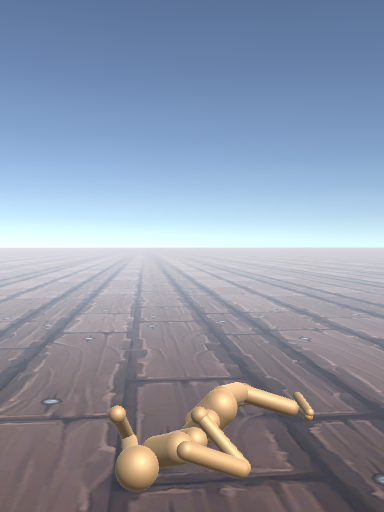}
      \end{subfigure}
      \begin{subfigure}{\figwidth}
        \includegraphics[width=\linewidth]{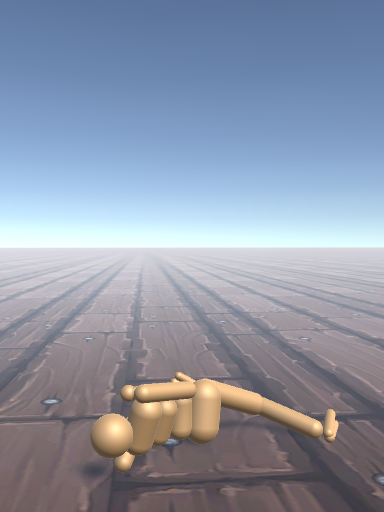}
      \end{subfigure}
      \begin{subfigure}{\figwidth}
        \includegraphics[width=\linewidth]{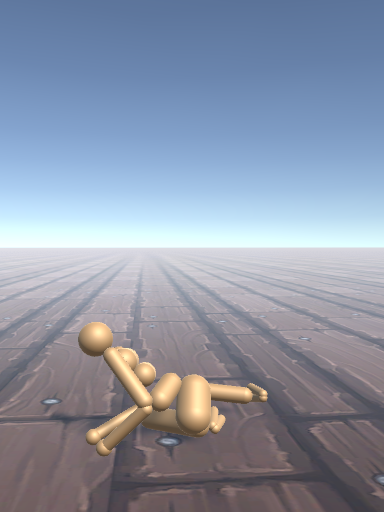}
      \end{subfigure}
      \begin{subfigure}{\figwidth}
        \includegraphics[width=\linewidth]{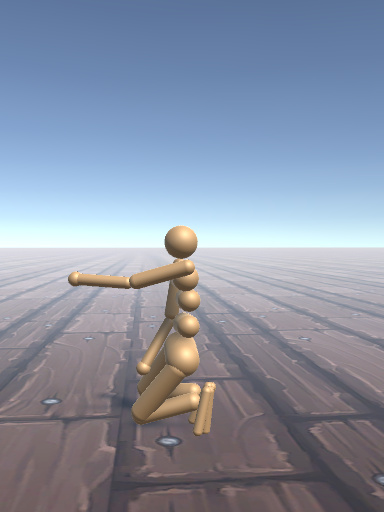}
      \end{subfigure}
      \begin{subfigure}{\figwidth}
        \includegraphics[width=\linewidth]{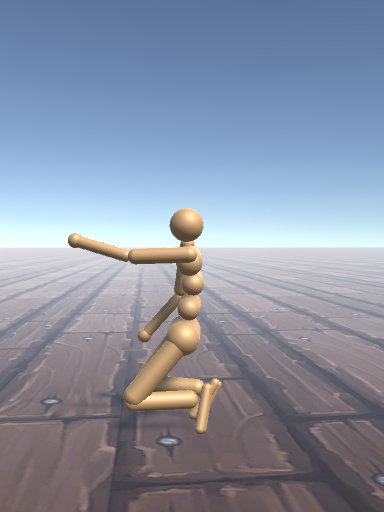}
      \end{subfigure}
      \begin{subfigure}{\figwidth}
        \includegraphics[width=\linewidth]{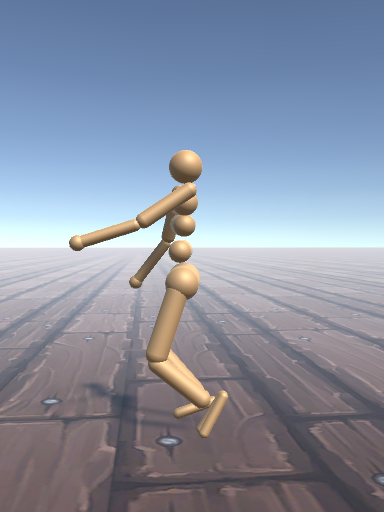}
      \end{subfigure}
      \begin{subfigure}{\figwidth}
        \includegraphics[width=\linewidth]{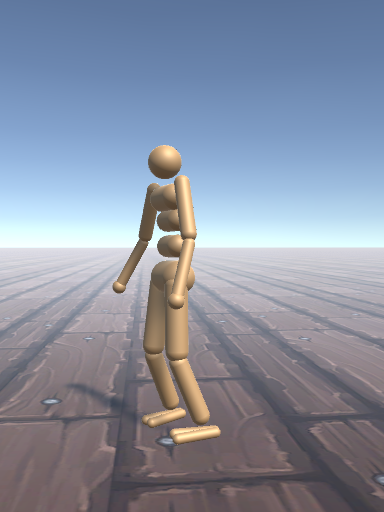}
      \end{subfigure}
      
      \caption{Controller D}
  \end{subfigure}

  \caption{Get-up motions of four controllers. Each 8 figures in a row show the get-up motion in one episode from the rag-doll fall to the standing phase. Two runs starting from supine and prone positions are shown for each controller.}
  \label{fig:four_controllers}
  
\end{figure}

\begin{figure}[h!]
  \def\figwidth{0.115\linewidth}
  \centering
  \begin{subfigure}{\linewidth}
        \begin{subfigure}{\figwidth}
          \includegraphics[width=\linewidth]{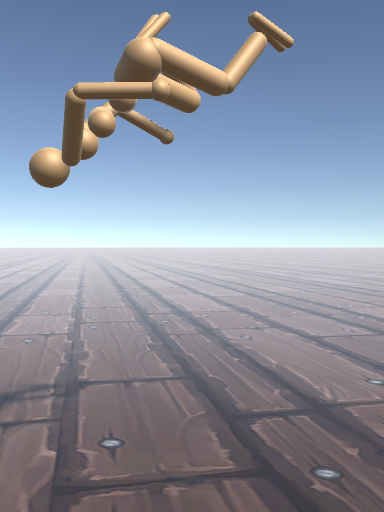}
        \end{subfigure}
        \begin{subfigure}{\figwidth}
          \includegraphics[width=\linewidth]{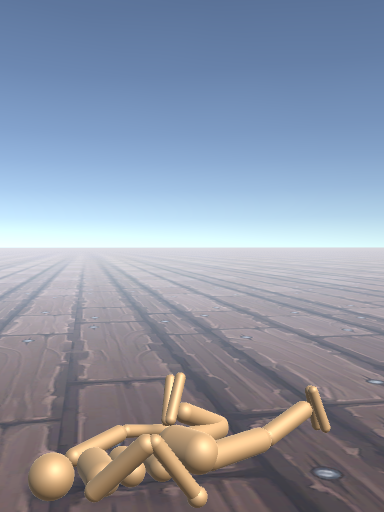}
        \end{subfigure}
        \begin{subfigure}{\figwidth}
          \includegraphics[width=\linewidth]{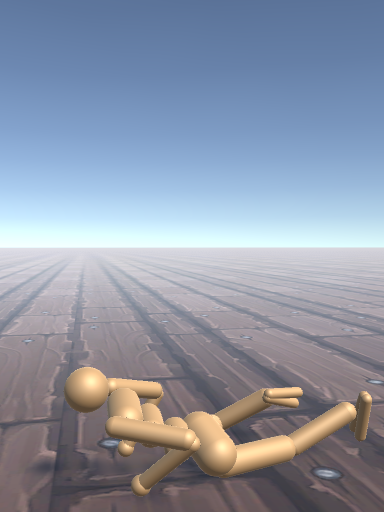}
        \end{subfigure}
        \begin{subfigure}{\figwidth}
          \includegraphics[width=\linewidth]{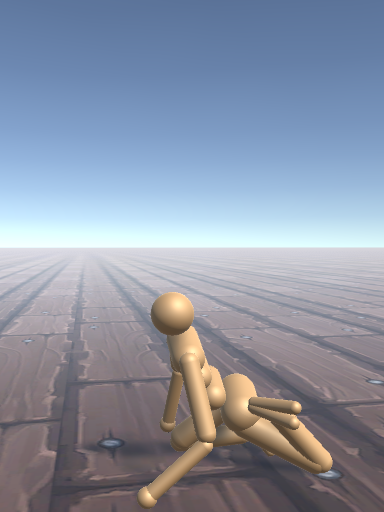}
        \end{subfigure}
        \begin{subfigure}{\figwidth}
          \includegraphics[width=\linewidth]{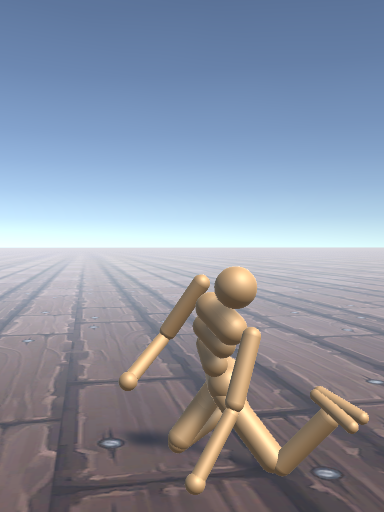}
        \end{subfigure}
        \begin{subfigure}{\figwidth}
          \includegraphics[width=\linewidth]{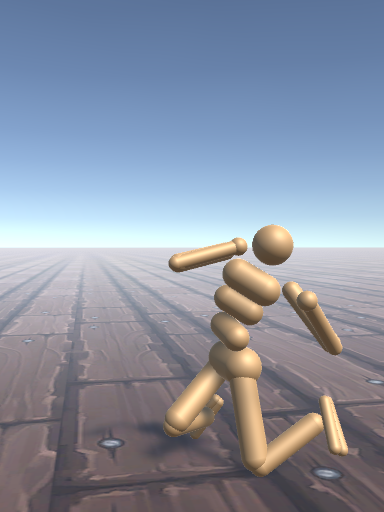}
        \end{subfigure}
        \begin{subfigure}{\figwidth}
          \includegraphics[width=\linewidth]{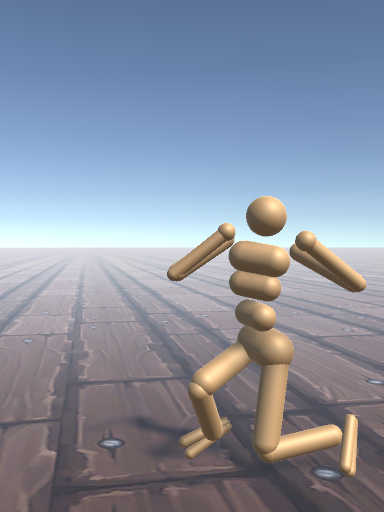}
        \end{subfigure}
        \begin{subfigure}{\figwidth}
          \includegraphics[width=\linewidth]{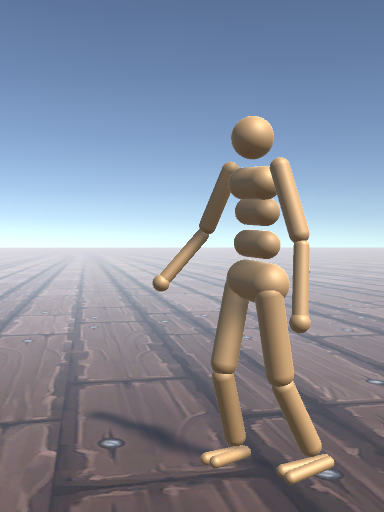}
        \end{subfigure}
  \end{subfigure}
  
  \begin{subfigure}{\linewidth}
      \begin{subfigure}{\figwidth}
        \includegraphics[width=\linewidth]{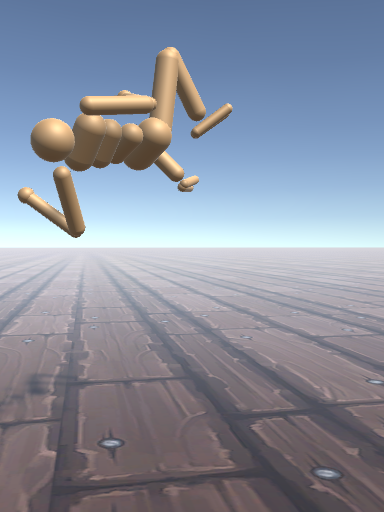}
      \end{subfigure}
      \begin{subfigure}{\figwidth}
        \includegraphics[width=\linewidth]{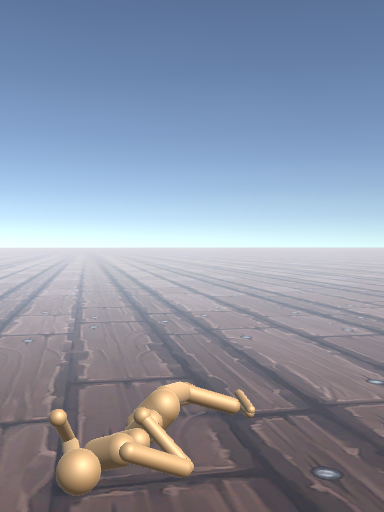}
      \end{subfigure}
      \begin{subfigure}{\figwidth}
        \includegraphics[width=\linewidth]{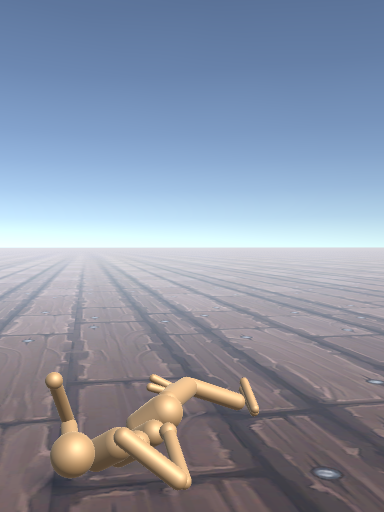}
      \end{subfigure}
      \begin{subfigure}{\figwidth}
        \includegraphics[width=\linewidth]{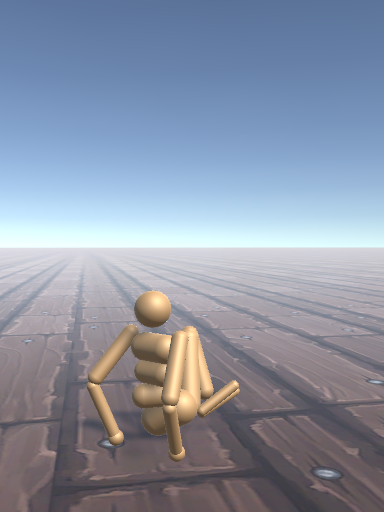}
      \end{subfigure}
      \begin{subfigure}{\figwidth}
        \includegraphics[width=\linewidth]{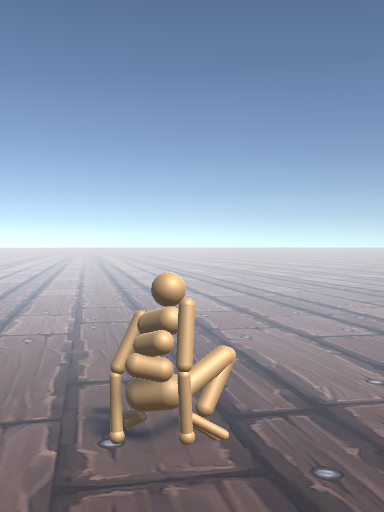}
      \end{subfigure}
      \begin{subfigure}{\figwidth}
        \includegraphics[width=\linewidth]{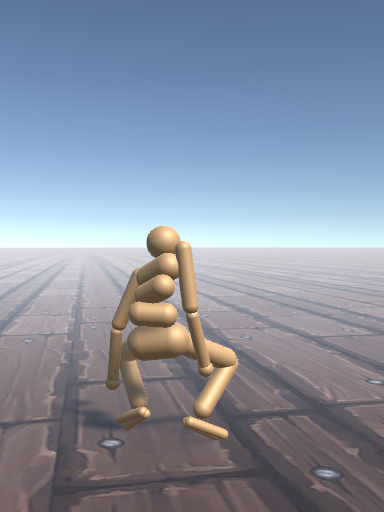}
      \end{subfigure}
      \begin{subfigure}{\figwidth}
        \includegraphics[width=\linewidth]{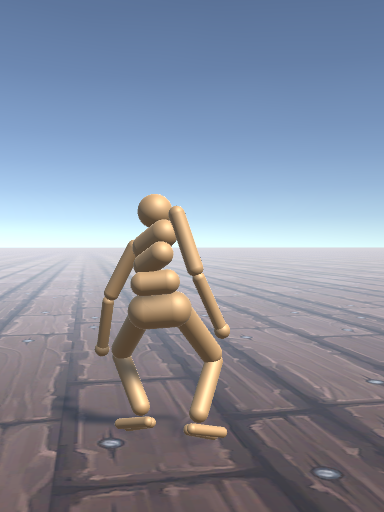}
      \end{subfigure}
      \begin{subfigure}{\figwidth}
        \includegraphics[width=\linewidth]{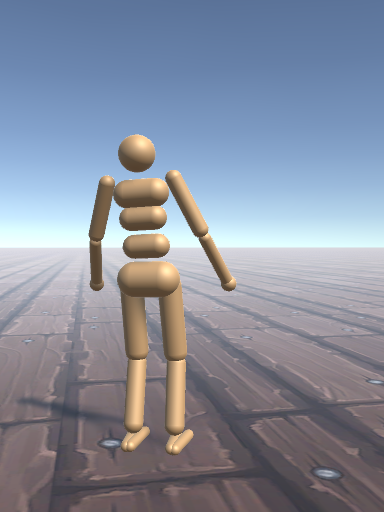}
      \end{subfigure}
  \end{subfigure}
  \caption{One get-up controller adopting different strategies starting from the supine and prone positions.}
  \label{fig:different_strategies}
\end{figure}

\begin{figure*}
    \centering
    \begin{subfigure}{0.23\linewidth}
        \includegraphics[width=\linewidth]{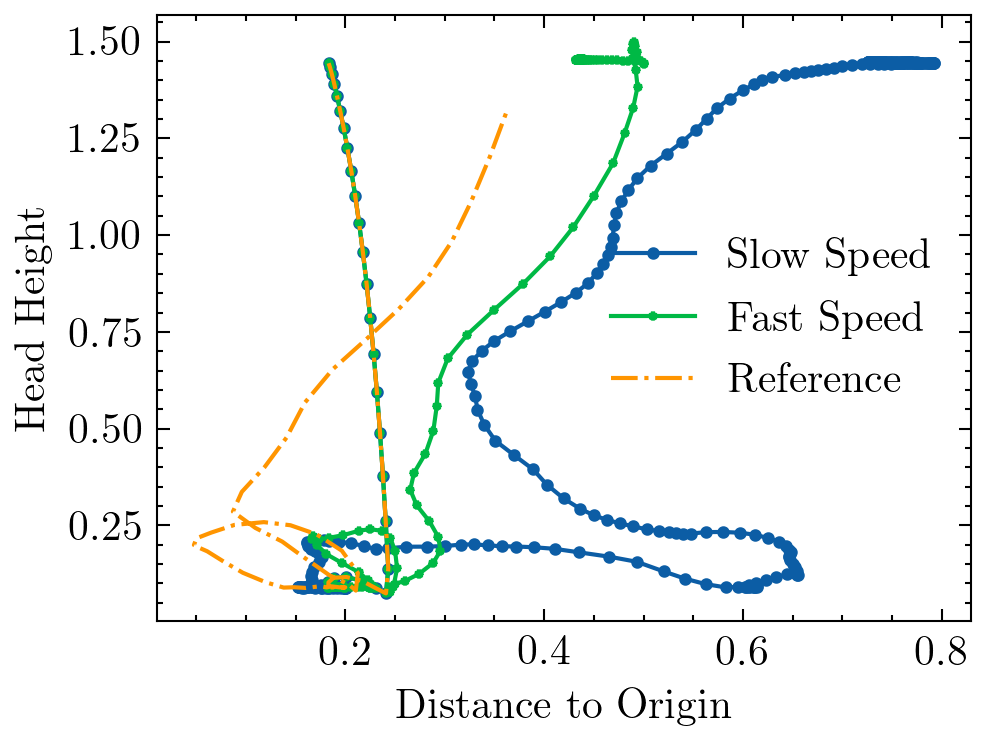}
        \caption{Controller A}
    \end{subfigure}
    \begin{subfigure}{0.23\linewidth}
        \includegraphics[width=\linewidth]{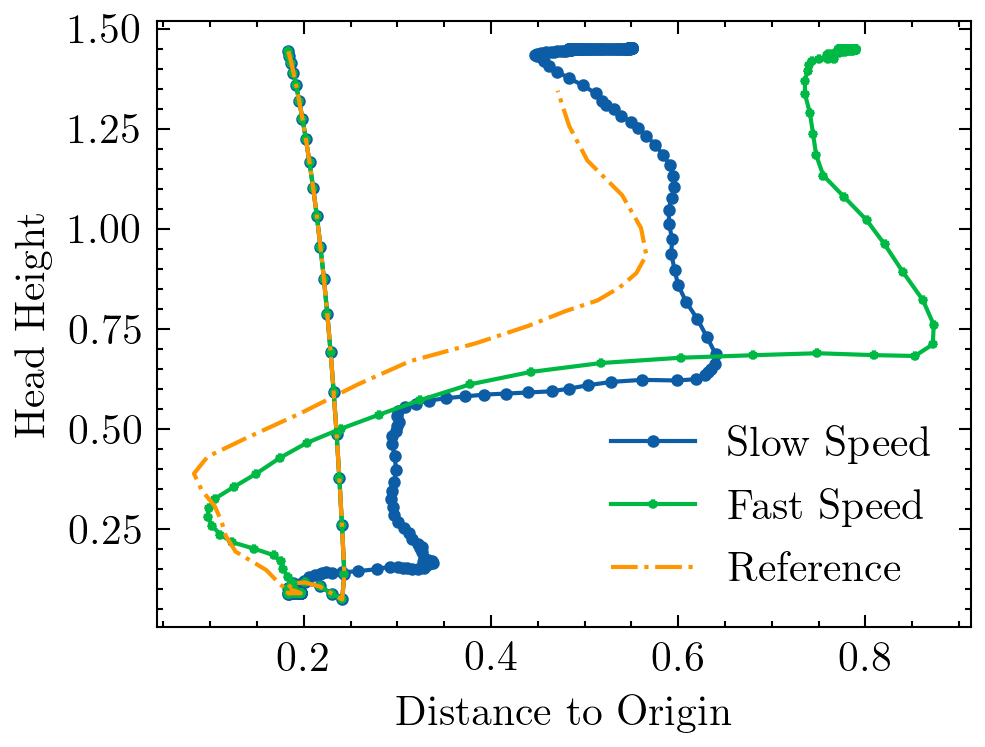}
        \caption{Controller B}
    \end{subfigure}
    \begin{subfigure}{0.23\linewidth}
        \includegraphics[width=\linewidth]{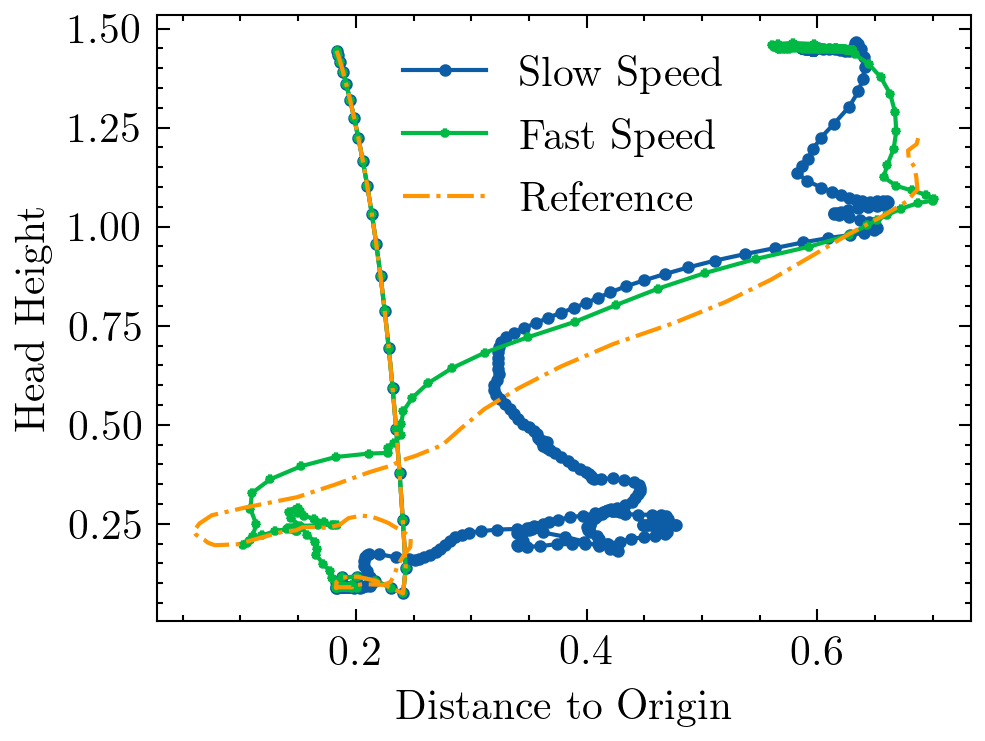}
        \caption{Controller C}
    \end{subfigure}
    \begin{subfigure}{0.23\linewidth}
        \includegraphics[width=\linewidth]{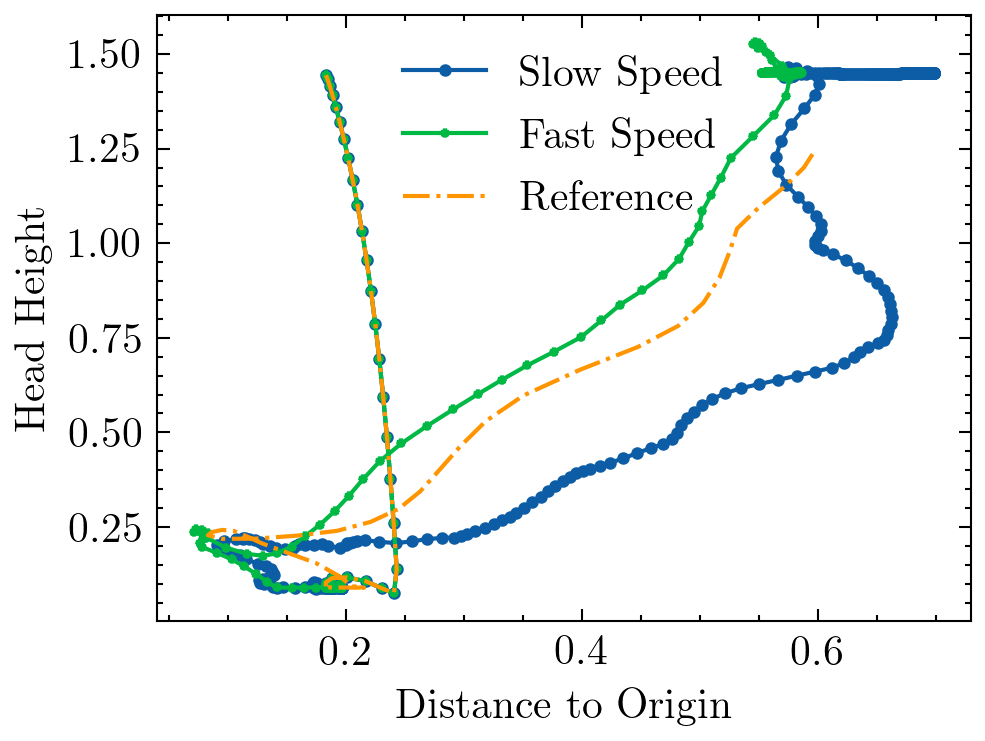}
        \caption{Controller D}
    \end{subfigure}

    \caption{Head trajectories in the lateral space of the character. Each plot includes the head trajectory in the lateral view for slow get-up, fast get-up and reference motion. There are noticeable differences between the three trajectories. We choose $\kappa = 0.25, 0.75$ for the slow and fast trajectories respectively. Each marker in the path represents one control timestep.}
    \label{fig:head_trajectories}
\end{figure*}

\begin{figure}
    \centering
    \begin{subfigure}{0.49\linewidth}
        \includegraphics[width=\linewidth]{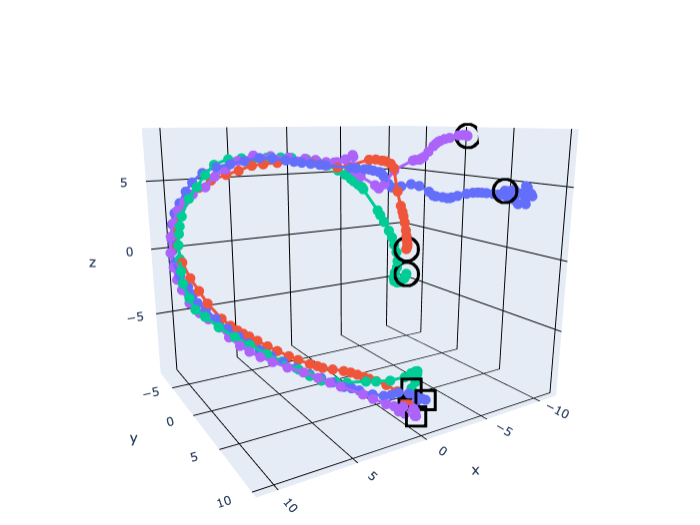}
        \caption{Different Initial States}
        \label{fig:same_controller}
    \end{subfigure}
    \begin{subfigure}{0.49\linewidth}
        \includegraphics[width=\linewidth]{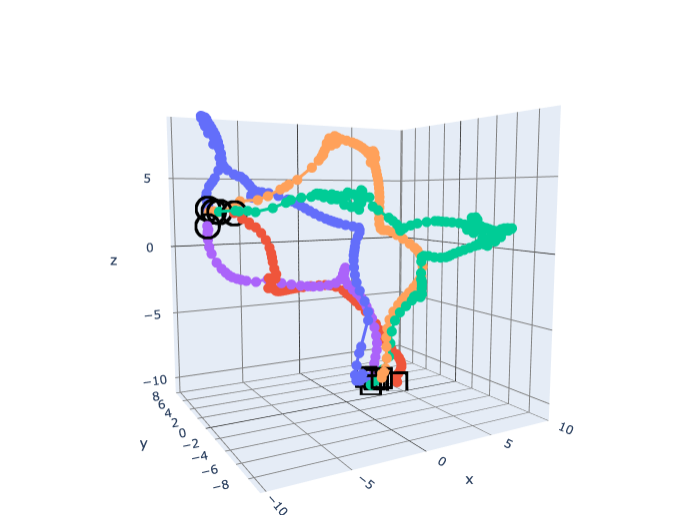}
        \caption{Different Controllers}
        \label{fig:different_controllers}
    \end{subfigure}\\
    \begin{subfigure}{0.49\linewidth}
        \includegraphics[width=\linewidth]{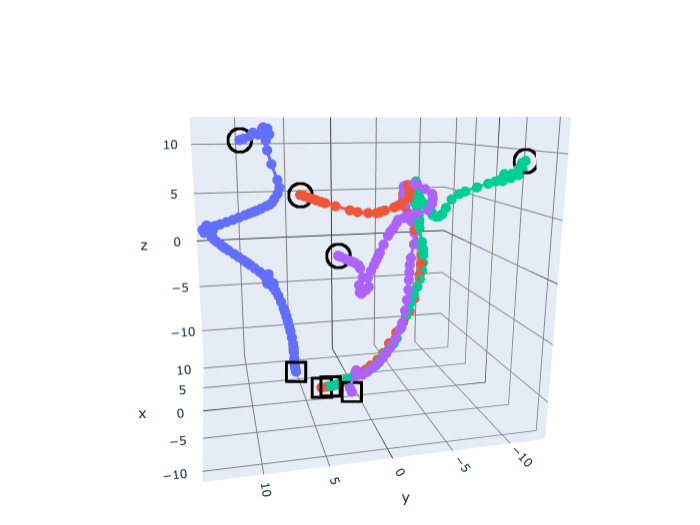}
        \caption{Different Controllers}
        \label{fig:different_strategies_tsne}
    \end{subfigure}
    
    \caption{t-SNE plots of trajectories. Fig.~a) shows four trajectories from the same controller starting from four different initial states. Fig.~b) shows five trajectories from different controllers starting from the same initial state. Fig.~c) shows four trajectories from the controller using different strategies when starting from supine and prone positions. The starting states are circled while the final states are squared.}
    \label{fig:tsne_plots}
\end{figure}

We demonstrate the learning curves for both fixed torque limits and strong-to-weak curriculum in Fig.~\ref{fig:learning_curve} and the value of torque limits with the curriculum in Fig.~\ref{fig:torque_limit_curve}. Fig.~\ref{fig:learning_curve} verifies the importance of proper torque limits for exploring a coarse solution mode. When the torque limit is fixed at $60\%$, $50\%$ and $40\%$ of the default torque limit~$\mathcal{T}$ respectively, the controller is likely to fail to discover any get-up solution mode. By employing the strong-to-weak curriculum, the agent first learns a solution mode and then refines the motion while adapting to the decreasing torque limits. As shown in Fig.~\ref{fig:torque_limit_curve}, the final torque limit with curriculum drops to below 60\% of the default torque limit~$\mathcal{T}$ on average. The strong-to-weak curriculum achieves comparable learning speed as the full strength model but produces a more natural get-up motion.

The strong-to-weak curriculum is terminated according to the rule described in Sec.~\ref{sec: torque_limit}. We find that the DRL policy tends to first adopt a fixed solution mode first and then refines it. Launching experiments with different seeding functions usually yields diverse get-up styles. Therefore, those get-up motions will end up with different final torque limits. As a subjective observation, we find the final torque limits to be correlated to the naturalness of the get-up motions. In our experiments, get-up strategies ending with torque limits ranging from 40$\%$ to 60$\%$ are usually perceived as more natural than those terminating with high torque limits~(above ~70$\%$). We include the get-up motions of several \piweak in the supplementary video. 

\subsection{Slow Get-up Motion}

We next show results at medium speed~($\kappa=0.5$), from the initial rag-doll fall to getting up from the ground and finally remaining to standing. Fig.~\ref{fig:four_controllers} shows four different controllers labelled as A, B, C and D starting from two initial states. Each controller either prefers to get up from the supine position or the prone position. If a controller prefers to get up in a supine position but starts from a prone position, the character commonly first rolls over, then gets up, and vice versa. However, exceptions also exist that attempt to get up from supine and prone positions using different strategies, as shown in Fig.~\ref{fig:different_strategies}. 

Our controller can produce get-up motions with different speeds by adjusting the retiming coefficient~$\kappa$. To analyse the behavior of controllers running at different speeds, we plot the trajectory of the head in the lateral space. Fig.~\ref{fig:head_trajectories} plots the head height versus the distance to the origin projected to the $xy$ plane for each controller we showed in Fig.~\ref{fig:four_controllers}. The slow and fast
trajectories share a similar path with the reference trajectory but
make their adaptions to accomplish the tasks, which indicates the necessity to learn a physics-based controller \pislow rather than 
merely being an identical-but-slower copy of the reference motion~$\tau_{\mathrm{fast}}$.


To better understand the structure and diversity of the learned get-up strategies, we project features of the state to a 3D space by t-SNE, and generate plots of the trajectory in 3D space. More details regarding the t-SNE implementation are included in App.~\ref{sec:tsne_implementation}. Fig.~\ref{fig:same_controller} shows the get-up trajectories for a given controller for four different initial states. These begin at different points in the embedded space and then merge to a single trunk because the given controller tends to adopt the same strategy to get up, and eventually arrive at the region representing the standing pose. Fig.~\ref{fig:different_controllers} reveals the differences across multiple controllers starting from an identical initial state. The projected trajectories begin with the same point after the rag-doll stage, then diverge to different paths to get up from the ground, and finally merge to the standing pose. Fig.~\ref{fig:different_strategies_tsne} shows the t-SNE trajectory plot for the controller adopting different strategies in supine and prone positions. Trajectories starting from supine and prone positions take different paths to the standing region in the embedded space.

\subsection{Paused Get-up Motion}

\begin{figure}
  \def\figwidth{0.233\linewidth}
  \centering
  \begin{subfigure}{\figwidth}
    \includegraphics[width=\linewidth]{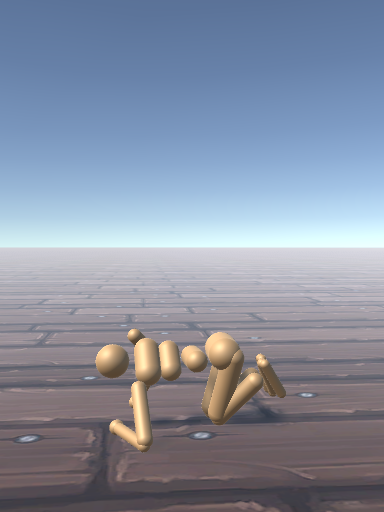}
    \caption{Controller A}
  \end{subfigure}
  \begin{subfigure}{\figwidth}
    \includegraphics[width=\linewidth]{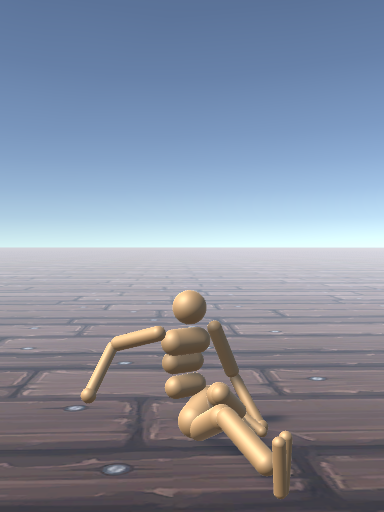}
    \caption{Controller B}
  \end{subfigure}
  \begin{subfigure}{\figwidth}
    \includegraphics[width=\linewidth]{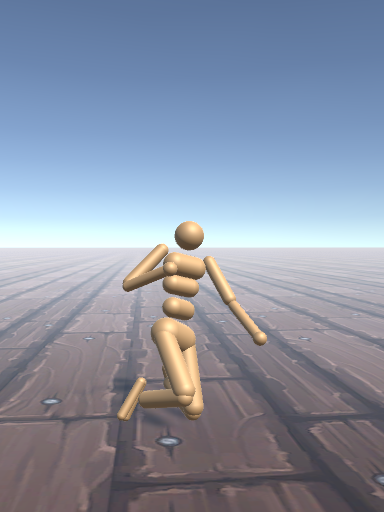}
    \caption{Controller C}
  \end{subfigure}
  \begin{subfigure}{\figwidth}
    \includegraphics[width=\linewidth]{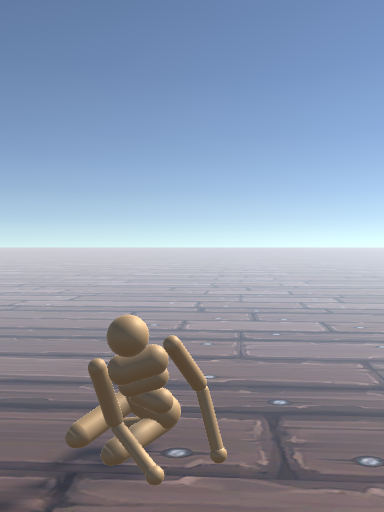}
    \caption{Controller D}
  \end{subfigure}
  \caption{Paused get-up states. Each picture shows one state that the get-up motion can be paused at.}
  \label{fig:paused_states}
\end{figure}

As the slow get-up policy \pislow is conditioned on two future poses, we can manipulate the reference trajectory \taufast in various ways other than uniform retiming. One idea is to repeat one specific state~$\hat{s}'$ in the reference trajectory multiple times such that the policy aims to reach the repeated pose~$\Tilde{q}$ first, then maintains the pose for a while, and finally continues the rest of the get-up motion. This setting creates a get-up motion paused at the repeated state. Without a future pose conditioned policy, such paused motion is nearly impossible to achieve with a purely state-indexed policy. As a result, our motion can be paused and continued in many statically stable states as shown in Fig.~\ref{fig:paused_states}, although the character loses balance when asked to pause in more dynamical states. We include the get-up motions with pauses in the supplementary video. In general, we find that the generated get-up motions are usually more statically stable at the beginning and become dynamic and less stable near the end of the get-up.

\subsection{Get-up Motion for Humanoid Variants}

\begin{figure}
  \def\figwidth{0.115\linewidth}
  \centering
  \begin{subfigure}{\linewidth}
      \begin{subfigure}{\figwidth}
        \includegraphics[width=\linewidth]{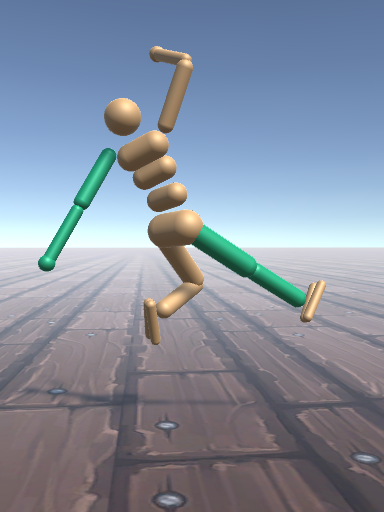}
      \end{subfigure}
      \begin{subfigure}{\figwidth}
        \includegraphics[width=\linewidth]{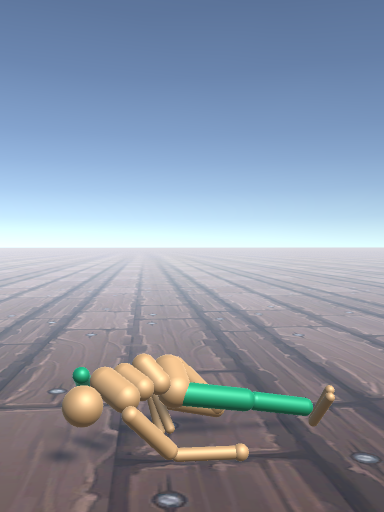}
      \end{subfigure}
      \begin{subfigure}{\figwidth}
        \includegraphics[width=\linewidth]{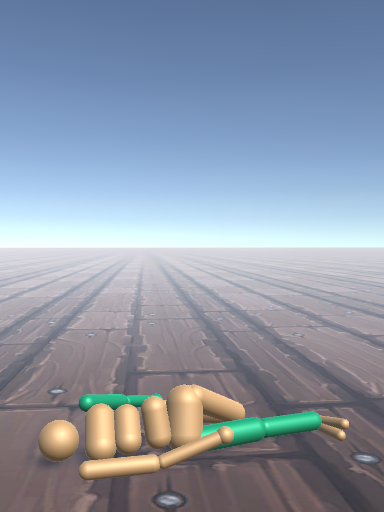}
      \end{subfigure}
      \begin{subfigure}{\figwidth}
        \includegraphics[width=\linewidth]{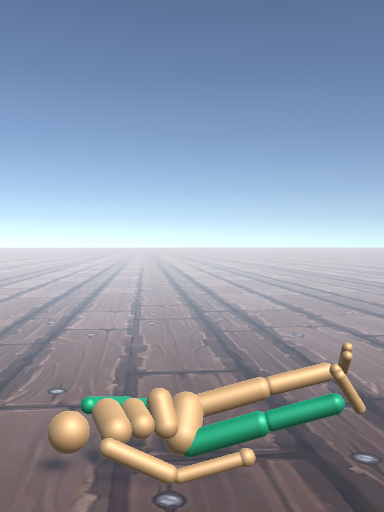}
      \end{subfigure}
      \begin{subfigure}{\figwidth}
        \includegraphics[width=\linewidth]{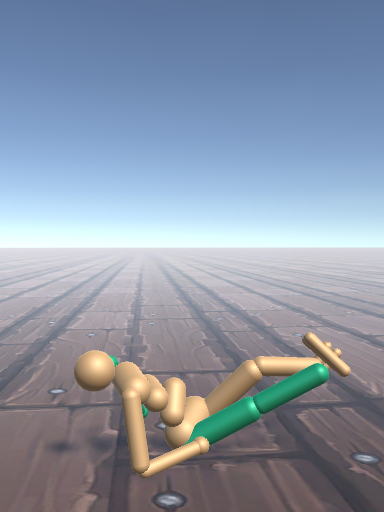}
      \end{subfigure}
      \begin{subfigure}{\figwidth}
        \includegraphics[width=\linewidth]{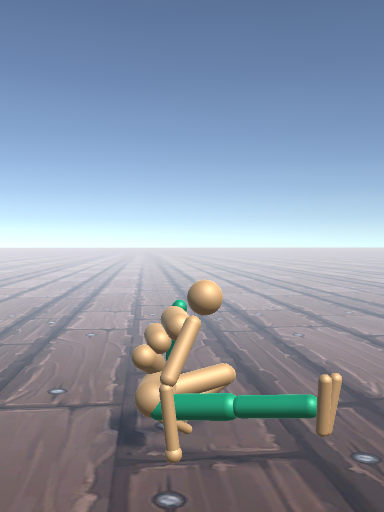}
      \end{subfigure}
      \begin{subfigure}{\figwidth}
        \includegraphics[width=\linewidth]{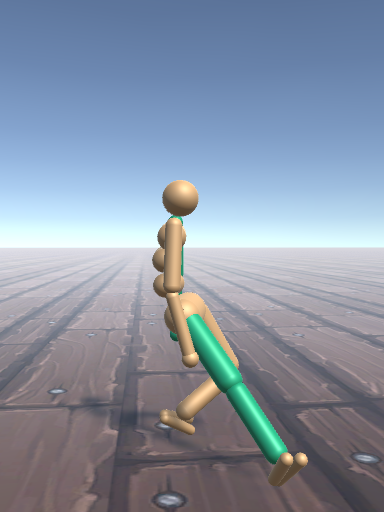}
      \end{subfigure}
      \begin{subfigure}{\figwidth}
        \includegraphics[width=\linewidth]{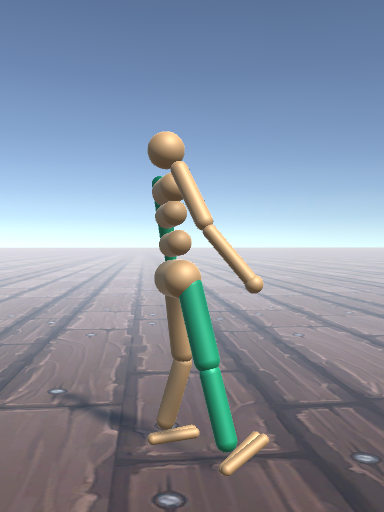}
      \end{subfigure}
  \end{subfigure}
  
  \begin{subfigure}{\linewidth}
      \begin{subfigure}{\figwidth}
        \includegraphics[width=\linewidth]{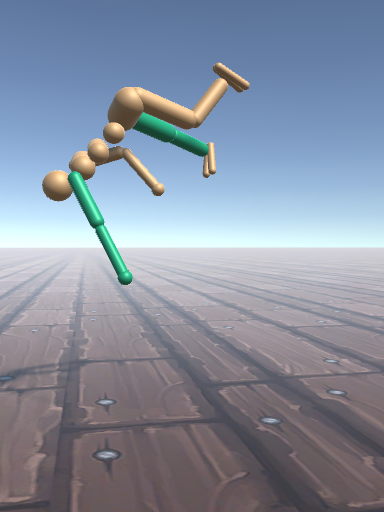}
      \end{subfigure}
      \begin{subfigure}{\figwidth}
        \includegraphics[width=\linewidth]{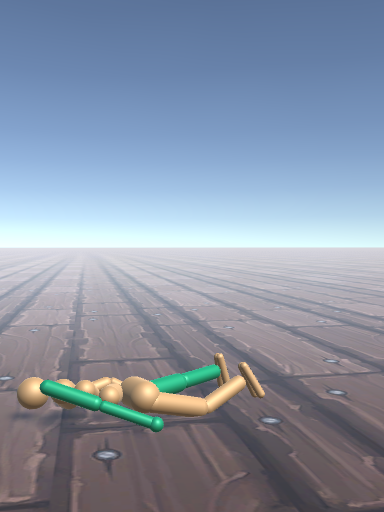}
      \end{subfigure}
      \begin{subfigure}{\figwidth}
        \includegraphics[width=\linewidth]{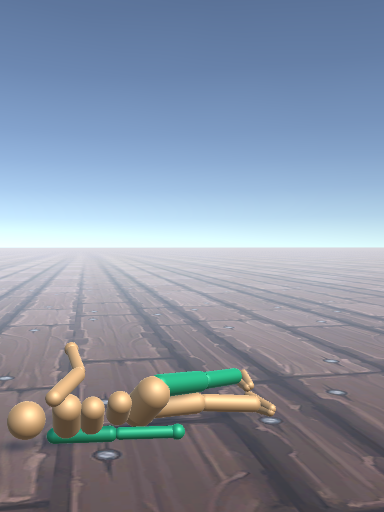}
      \end{subfigure}
      \begin{subfigure}{\figwidth}
        \includegraphics[width=\linewidth]{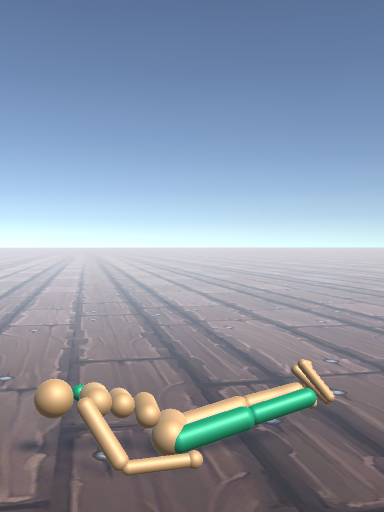}
      \end{subfigure}
      \begin{subfigure}{\figwidth}
        \includegraphics[width=\linewidth]{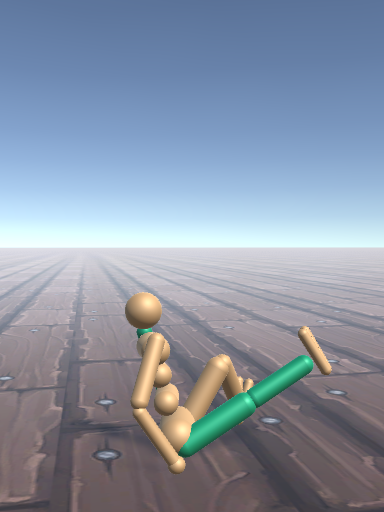}
      \end{subfigure}
      \begin{subfigure}{\figwidth}
        \includegraphics[width=\linewidth]{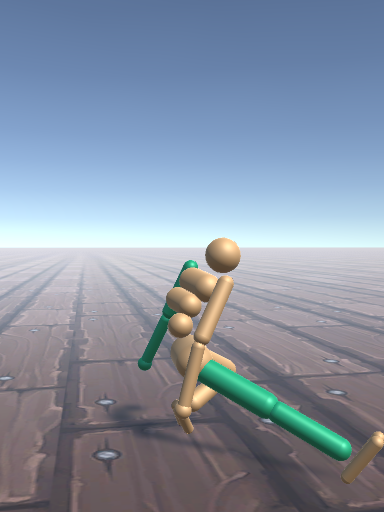}
      \end{subfigure}
      \begin{subfigure}{\figwidth}
        \includegraphics[width=\linewidth]{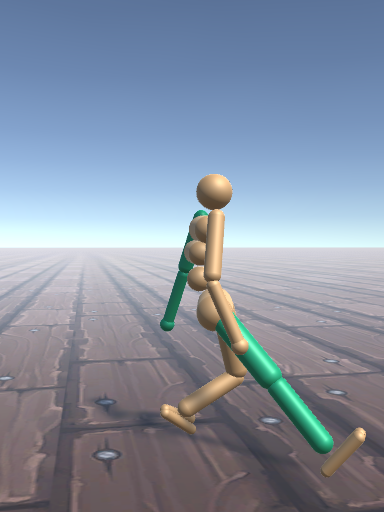}
      \end{subfigure}
      \begin{subfigure}{\figwidth}
        \includegraphics[width=\linewidth]{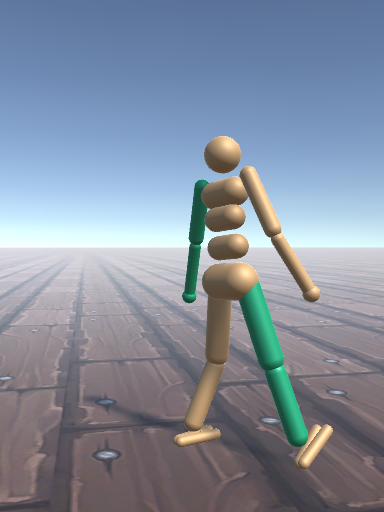}
      \end{subfigure}
  \end{subfigure}
  
  \caption{Get-up motions for humanoid with a leg and an arm in a cast. The left elbow joint and right knee joint are locked throughout the motion. The corresponding limbs are rendered in green. Each row of images shows the get-up motion from either the supine position or the prone position.}
  \label{fig:disabled_humanoid}
\end{figure}

\begin{figure}
  \def\figwidth{0.115\linewidth}
  \centering
  \begin{subfigure}{\linewidth}
      \begin{subfigure}{\figwidth}
        \includegraphics[width=\linewidth]{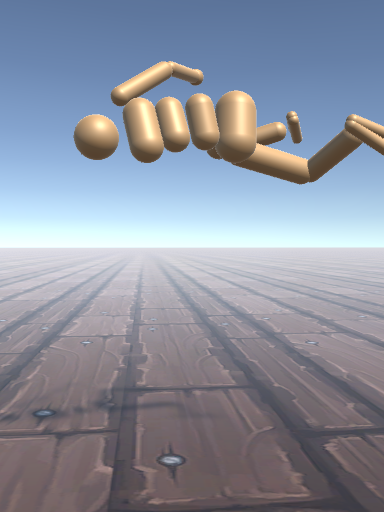}
      \end{subfigure}
      \begin{subfigure}{\figwidth}
        \includegraphics[width=\linewidth]{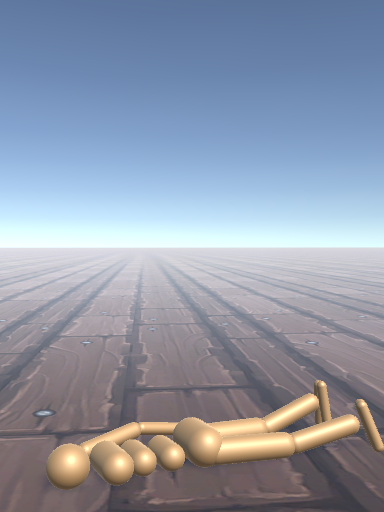}
      \end{subfigure}
      \begin{subfigure}{\figwidth}
        \includegraphics[width=\linewidth]{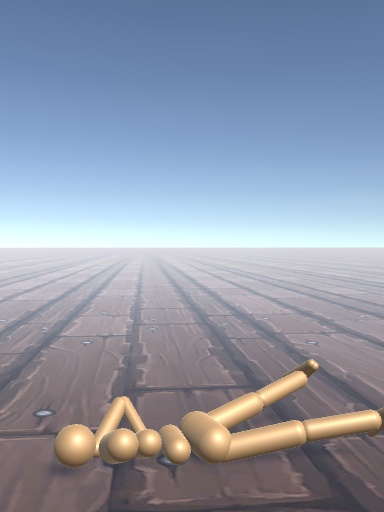}
      \end{subfigure}
      \begin{subfigure}{\figwidth}
        \includegraphics[width=\linewidth]{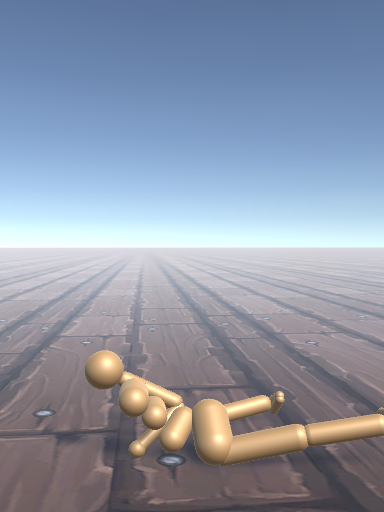}
      \end{subfigure}
      \begin{subfigure}{\figwidth}
        \includegraphics[width=\linewidth]{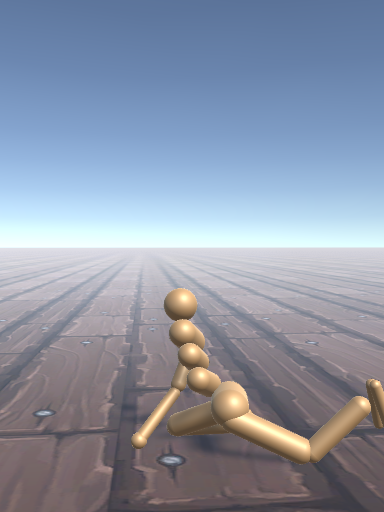}
      \end{subfigure}
      \begin{subfigure}{\figwidth}
        \includegraphics[width=\linewidth]{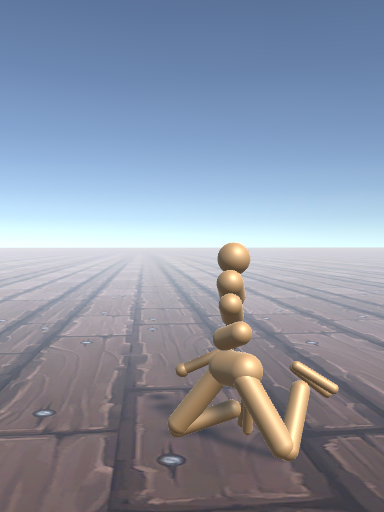}
      \end{subfigure}
      \begin{subfigure}{\figwidth}
        \includegraphics[width=\linewidth]{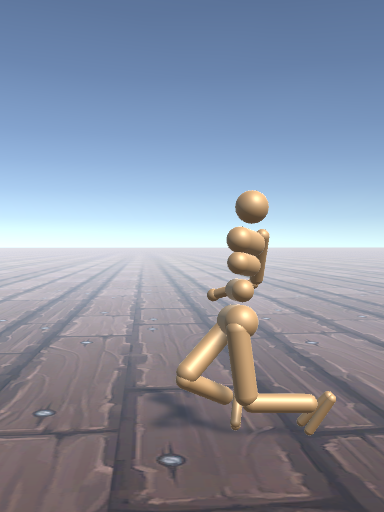}
      \end{subfigure}
      \begin{subfigure}{\figwidth}
        \includegraphics[width=\linewidth]{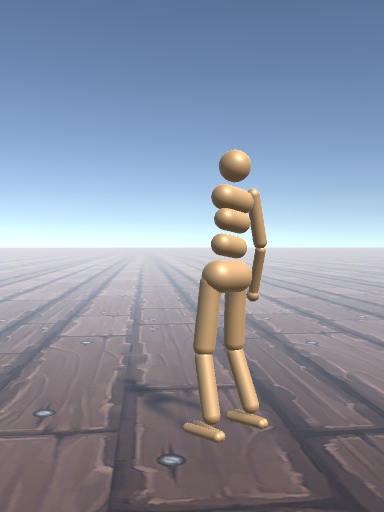}
      \end{subfigure}
  \end{subfigure}
  
  \begin{subfigure}{\linewidth}
      \begin{subfigure}{\figwidth}
        \includegraphics[width=\linewidth]{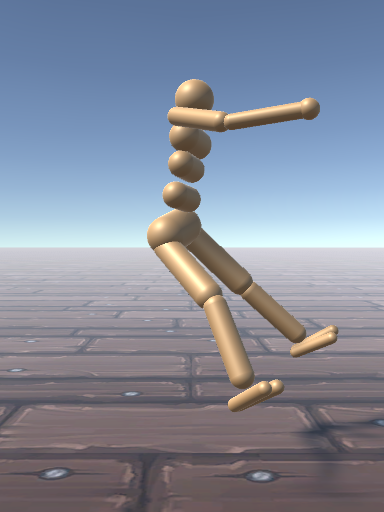}
      \end{subfigure}
      \begin{subfigure}{\figwidth}
        \includegraphics[width=\linewidth]{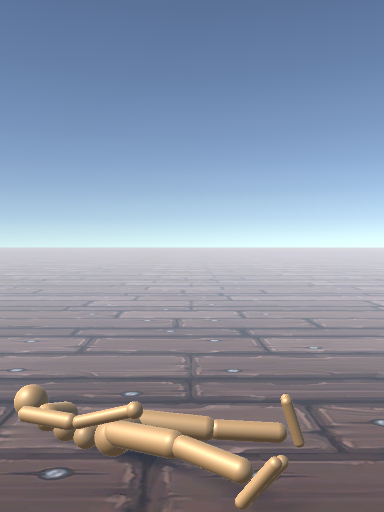}
      \end{subfigure}
      \begin{subfigure}{\figwidth}
        \includegraphics[width=\linewidth]{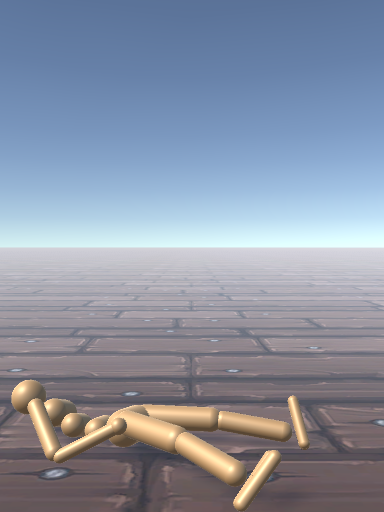}
      \end{subfigure}
      \begin{subfigure}{\figwidth}
        \includegraphics[width=\linewidth]{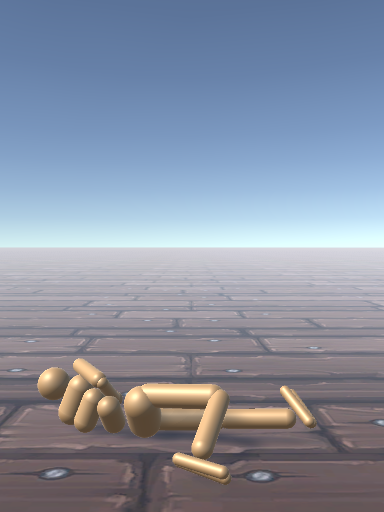}
      \end{subfigure}
      \begin{subfigure}{\figwidth}
        \includegraphics[width=\linewidth]{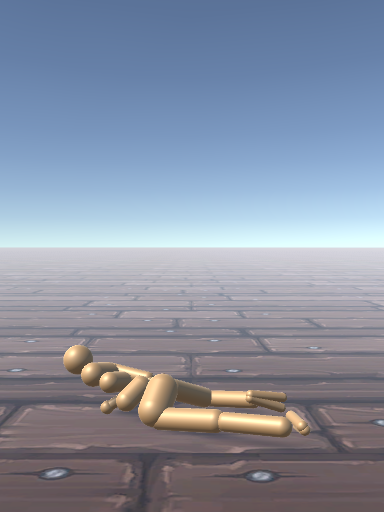}
      \end{subfigure}
      \begin{subfigure}{\figwidth}
        \includegraphics[width=\linewidth]{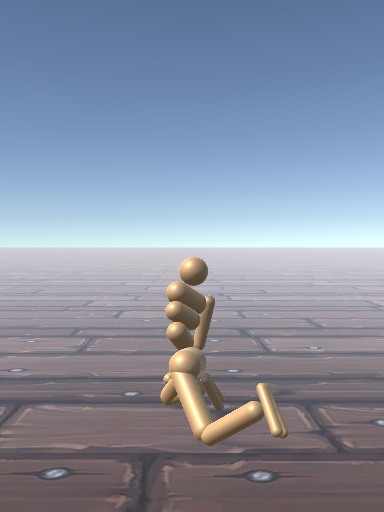}
      \end{subfigure}
      \begin{subfigure}{\figwidth}
        \includegraphics[width=\linewidth]{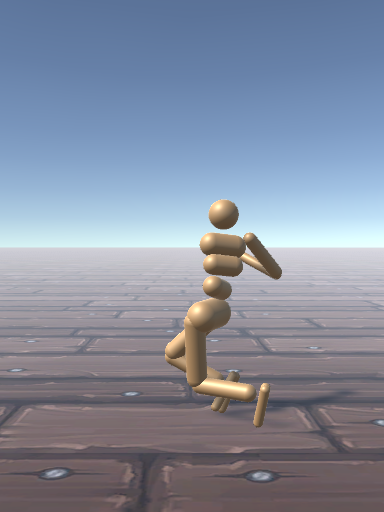}
      \end{subfigure}
      \begin{subfigure}{\figwidth}
        \includegraphics[width=\linewidth]{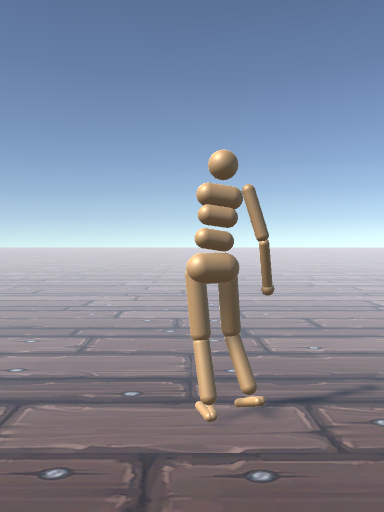}
      \end{subfigure}
  \end{subfigure}
  
  \caption{Get-up motions for humanoid with missing left arm. Each row of images shows the get-up motion from either the supine position or the prone position.}
  \label{fig:noarm_humanoid}
\end{figure}

Following the same pipeline, we can generate get-up motions adapted to a humanoid character with a leg and an arm in a cast and a humanoid character with a missing arm. Fig.~\ref{fig:disabled_humanoid} and Fig.~\ref{fig:noarm_humanoid} illustrate the discovered get-up motions for those special characters respectively. The resulting motion is also demonstrated in the supplementary video. We show that these irregular humanoid models can still get up at various commanded speeds. Since two joints are removed for the character with limbs in casts, it adopts a get-up strategy that relies on the remaining limbs to gain momentum, while using the limbs in casts for balance at certain stages. The policy developed with the character with a missing arm attempts to get up by pushing against the ground using one arm only. These results show that our pipeline is not restricted to a specific model but can be applied to situations where motion capture data is hard to obtain.

\section{Ablation Studies}

We conduct multiple ablation studies to investigate the role of several components in our system, and explore alternative methods to learn weak and slow get-up motions via reward engineering. We experiment with removing the strong-to-weak curriculum and slow get-up imitation respectively. As a result, we generally observe that the resulting motions are of lower quality. 
We note that careful reward design itself is not sufficient to learn a weak and slow get-up motion. Relevant videos are included in the supplementary material.

\paragraph{High Strength Get-up Motion.}
We first illustrate that training with high-strength characters tends to find an unnatural and highly-dynamic get-up motion. We train the initial policy \pistrong without any modification on the torque limit. 
The results are best observed in the supplemental video.
The strong-to-weak curriculum eliminates excessively aggressive and abrupt motions.


\paragraph{Weak Motions Without a Curriculum.}
As discussed earlier~(\S\ref*{sec:torque_limit_results}), starting the training with fixed low torque limits typically traps the policy in local minima and fails to find any suitable solution mode.

\paragraph{Slow Get-up Imitation Without Curriculum.}
We also evaluate an ablation where we skip the strong-to-weak curriculum and proceed directly to imitating retimed versions of a strong policy~$\pi_{\mathrm{strong}}$. We find that \pistrong provides excessively-dynamic get-up motions for \pislow to imitate. As a result, the character fails to get up at low speeds. Although \pislow sometimes succeeds in fast get-up tasks, the motion remains overly dynamic and awkward.

\paragraph{Weak and Slow Get-up Motion Alternatives.}
Instead of an explicit imitation objective, motion constraints can often be embedded in the reward function design. Adding an energy cost term has been proposed in \cite{xie2020allsteps,ma2021learning, fu2021minimizing} to improve motion quality. However, we find that adding an energy cost without the strong-to-weak curriculum has minimal effect on the get-up motion. In addition, we experiment with adding a reward term to penalize high joint velocities for learning slow get-up motions. However, such a reward design either has negligible effects on the get-up speeds with fewer weights on this term or leads to training instability with more weights on it. We further test the option of annealing the weights of the energy cost term and the velocity regularization term linearly. Nevertheless, the resulting motions are not clearly more statically stable and slowed down. We include the corresponding motions in the supplementary video.


\section{Conclusion}

We have presented a framework based on deep reinforcement learning to produce natural human get-up from the ground motions without recourse to motion capture data. The final learned policy can yield realistic get-up motions at different speeds and from arbitrary initial states. We first exploit the benefits of a high-strength character to discover a particular get-up strategy. The initial policy is then refined with a progressively weaker character to enhance motion quality. Lastly, our method learns an imitation controller to get up at much slower speeds, including pausing in intermediate statically-stable states. We visualize the diversity across different controllers and the behavior from different initial states.

Our method has a variety of remaining limitations, pointing to directions for future work.
Currently our method still requires a separate simulation using \piweak in order to generate the reference trajectory that is used to condition $\pi_{\mathrm{slow}}$. It should be possible to learn a single policy that is directly conditioned on the current state and $\kappa$, mainly via a distillation.  This would eliminate the need to store and use $\pi_{\mathrm{weak}}$.  We leave this as future work. 

Our learning framework can discover, in a tabula rasa fashion, diverse get-up motions, across different runs with different randomized policy initializations. However, there is currently no means to provide user control. We wish to explore various possible methods for adding control over the choice of get-up strategy, and more general control over the style.  One interesting strategy would be to learn a set of $N$ controllers, and then have a user specify their preference for the desired get-up strategy employed from different initial states, and to then reintegrate this into a single controller.

Humans need to get up from chairs, sofas, bathtubs, car seats, variable terrain, and a variety of other constrained situations, e.g., getting up while wearing skates or skis. Humans are extremely adept at finding good solutions to these problems. An exciting direction for future work will be to produce controllers that can generalize well to this broad range of circumstances.


\begin{acks}
We thank Sheldon Andrews, Anthony Frezzato, and Arsh Tangri for many useful discussions about the getup problem.
\end{acks}


\bibliographystyle{ACM-Reference-Format}
\bibliography{bibliography}
\clearpage

\appendix
{\Large \bf Supplemental Material for Learning to Get Up}
\section{Pseudocode}
\label{sec:pseudocode}

We provide the pseudocode of the Discover and the Weaker stages in Algorithm~\ref{alg:discover_and_weaker}, and that of the Slower stage in Algorithm~\ref{alg:slower}.

\begin{algorithm}
\SetAlgoNoLine
\KwIn{Torque limit multiplier $\beta$, Minimum steps $N_{min}$, Maximum steps $N_{max}$, Policy $\pi$, Threshold $\omega$}
\textit{// Initialize parameters}\;
$i \gets 0$\;
$N \gets 0$\;
$N_{c} \gets 0$\;
$\hat{\beta} \gets 1.0$\;
\texttt{Initialize initial state to $s$}\;
\For{step = 1, \dots, max\_steps}
{
    $a \sim \pi(s,\hat{\beta})$\;
    $(s',R, done) = \texttt{ForwardSimulation(s,a)}$\;
    \texttt{Save (s, a, R, s', done) in the replay buffer $D$}\;
    $s \gets s'$\;
    \texttt{Perform SAC update using data from $D$}\;
    $N_{c} = N_{c} + 1$\;
    \If{step \% test\_frequency == 0}{
        $R_{test} = \texttt{TestPolicy($\pi$)}$\;
        \textit{// Advance curriculum}\;
        \eIf{$R_{test} > \omega$}{ 
            $i = i + 1$\; 
            \textit{// Update termination criteria}\;
            $N = \texttt{Clip}(N_{c}, N\_min, N\_max)$\;
            $N_{c} = 0$\;
        }{
            \textit{// Terminate curriculum}\;
            \If{$N_{c} > N$}{
                \texttt{End Training}\;
            }
        }
    }
    \If{done}{
        \textit{// Sample the torque limit multiplier}\;
        $\hat{\beta} \sim \mathcal{N}(\beta^{i},\epsilon)$\;
        $s = \texttt{ResetEpisode()}$\;
    }
}
\caption{Discover and Weaker Stages}
\label{alg:discover_and_weaker}
\end{algorithm}

\begin{algorithm}
\SetAlgoNoLine
\KwIn{Weak Policy $\pi_{weak}$, Low speed limit $k_{low}$, High speed limit $k_{high}$}
$done \gets True$\;
\For{step = 1, \dots, max\_steps}
{
    \If{done}{
        $s = \texttt{ResetEpisode()}$\;
        $\tau_{fast} = \texttt{GenerateTrajectory(s, $\pi_{weak}$)}$\;
        $k \sim \mathcal{U}(k_{low}, k_{hight})$\;
        $\tau_{slow} = \texttt{Retime($\tau_{fast}$, k)}$\;
    }
    $a \sim \pi_{slow}(s,\tau_{slow})$\;
    $(s',R, done) = \texttt{ForwardSimulation(s,a)}$\;
    \texttt{Save (s, a, R, s', done) in the replay buffer $D$}\;
    $s \gets s'$\;
    \texttt{Perform SAC update to $\pi_{slow}$ using data from $D$}\;
}
\caption{Slower Stages}
\label{alg:slower}
\end{algorithm}
\section{Detailed Reward Design}
\label{sec:reward_design}

\begin{figure}[h]
  \centering
  \includegraphics[width=0.75\linewidth]{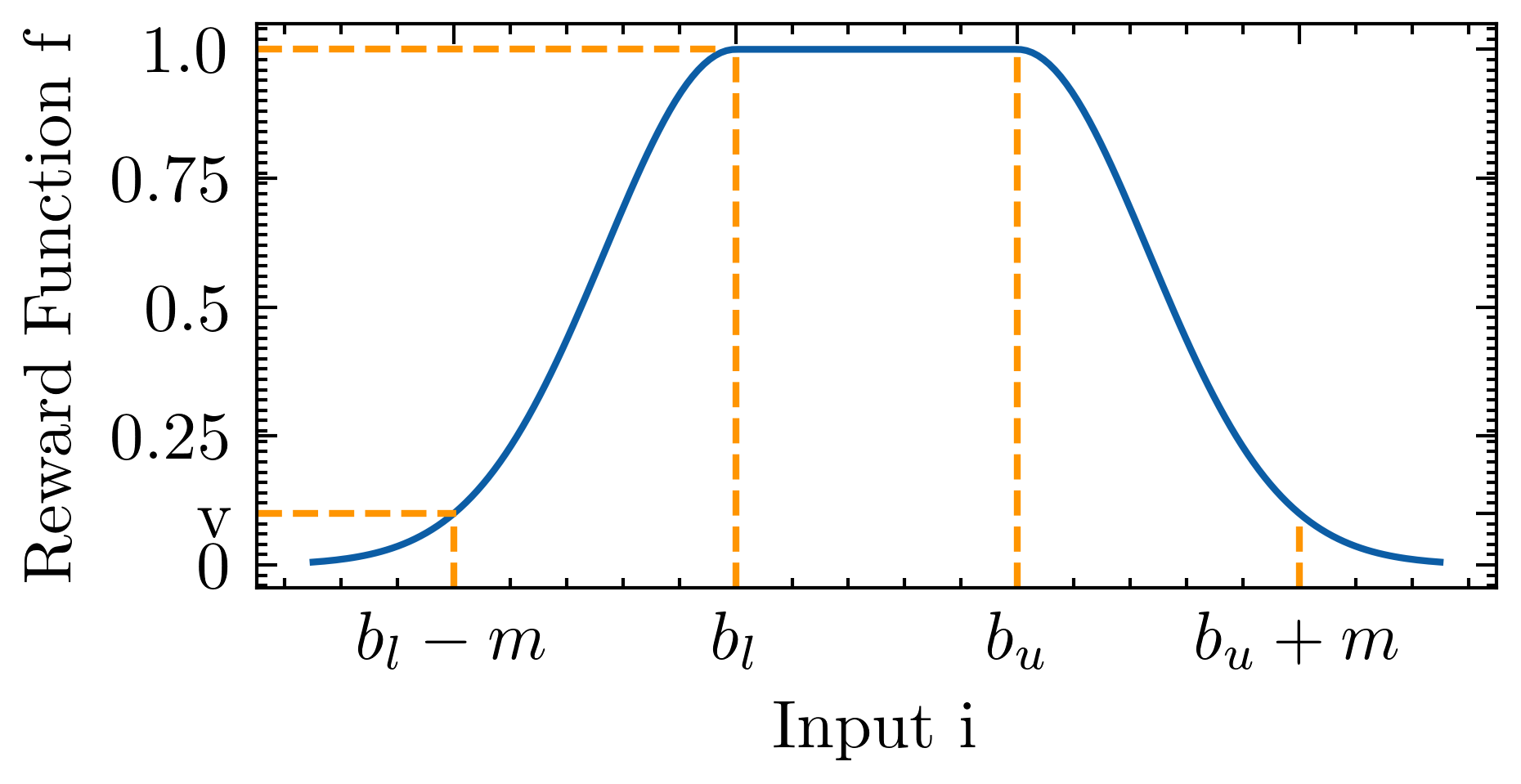}
  \caption{Reward Function $f(i,b,n,v)$. The general reward function is bounded in [0, 1] and specified by three parameters, bounds $b=[b_{l}, b_{u}]$, margin $v$ and ending value $v$.}
  \label{fig:reward_function}
\end{figure}

The overall reward function~$R(s_{t}, a_{t})$ is the product of multiple reward terms~$r$ between 0 and 1. As shown in Figure.~\ref{fig:reward_function}, each reward term~$r$ is calculated by a function of input value $i$ defined by three parameters bounds~$b$, margin~$m$ and value~$v$ as~$f(i,b,m,v)$. Bounds $b=[b_{l}, b_{u}]$ defines the region where the reward term is 1 if the input value $i$ is inside. The reward value will drop smoothly outside the bounds following a Gaussian curve until reaching value~$v$ at a distance of margin~$m$~\cite{tassa2020dmcontrol}.

\subsection{Exploring Initial Policies and its Low-energy Variants.}
One direct indication of the get-up behavior is the head height. Therefore, the agent will be rewarded $1$ once the head height is above a threshold. As the character gets up from the ground, the head height reward grows up to $1$ until reaching the threshold. We set the reward function as follows:
\begin{equation}
    r_{\mathrm{h}} = f(i=h_{\mathrm{head}},b=[1.55,inf],m=0.37,v=0.1).
\end{equation}

Additionally, the character is also encouraged to keep the torso vertically straight when getting up. We add a reward based on the vertical projection of the unit torso up vector, i.e. $z^{\mathrm{up}}_{\mathrm{torso}}$. The reward is only applied when the center of mass height $h_{com}$ is above $0.5m$ as:
\begin{equation}
    r_{straight} = 
    \begin{cases}
    f(i=z^{\mathrm{up}}_{\mathrm{torso}},b=[0.9,inf],m=1.9,&v=0.0),\\
    & \text{if } h_{\mathrm{com}} > 0.5\\
    \\
    1.0, & \text{otherwise}
    \end{cases}.
\end{equation}

The velocity is one way to distinguish a walking or running motion from a get-up motion. To prevent the agent from being rewarded for walking or running, we add a reward term constraining on the center of mass velocity projected on the horizontal plane~$v_{\mathrm{com}}^{\mathrm{xy}} = [v_{\mathrm{com}}^{\mathrm{x}},v_{\mathrm{com}}^{\mathrm{y}}]$:
\begin{equation}
    r_{v_{\mathrm{com}}^{\mathrm{xy}}} = \frac{1}{2}\sum_{v' \in v_{\mathrm{com}}^{\mathrm{xy}}}f(i=v',b=[-0.3,0.3],m=1.2,v=0.1),
\end{equation}
where we set the bounds parameter $b$ small to allow slow roll over motion.

We also observe that get-up motion with further split feet is perceived unnatural for humans. Thus, we constrain the distance between the feet by adding a penalty term. The penalty term encourages the feet to keep the euclidean feet distance projected to the $xy$ plane $d_{\mathrm{feet}}$ no more than roughly twice the shoulder width:
\begin{equation}
    r_{\mathrm{feet}} = f(i=d_{feet},b=[0,0.9],m=0.38,v=0).
\end{equation}

\subsection{Slow Get-up Motions}

For clarity and compactness, we denote the attributes in the reference trajectory \tauslow with an apostrophe as $(.)'$.

We first compare the center of mass height between the character and reference trajectory \tauslow at the current timestep by computing the distance $\Delta h_{\mathrm{com}} = h_{\mathrm{com}} - h_{\mathrm{com}}'$. The reward term can be formally expressed as:
\begin{equation}
    r_{\mathrm{com}} = f(i=\Delta h_{\mathrm{com}}, b=[0,0], m=0.5,v=0.1).
\end{equation}

Similarly, we also enforce the torso orientation vectors projected to the vertical axis between the character and reference trajectory to match. Therefore, we include a reward term on the vertical axis projection of the torso orientation vector $o_{\mathrm{torso}}=[x_{\mathrm{torso}}^{\mathrm{up}}, y_{\mathrm{torso}}^{\mathrm{up}}, z_{\mathrm{torso}}^{\mathrm{up}}]$ to the Cartesian coordinate:
\begin{equation}
    r_{\mathrm{ori}} = \prod_{o \in \Delta o}f(i=o, b=[-0.03, 0.03], m=0.6,v=0.3),
\end{equation}
where $\Delta o$ represents the difference between the orientation vectors, i.e., $\Delta o = o - o'$.

In addition, we notice that the tracking controller might exhibit unnatural swift stepping motion to gain balance during the get-up motion. This behavior is mainly caused by fast manipulating the hip joint motor along the y axis. To mitigate this issue, we add another regularization term on the angular joint velocities on both hip joints~$v_{\mathrm{hip}} = [v_{\mathrm{hip}}^{\mathrm{l}}, v_{\mathrm{hip}}^{\mathrm{r}}]$ to match the corresponding value in the reference trajectory:
\begin{equation}
    r_{\mathrm{hip}} = \frac{1}{2}\sum_{\Delta v \in v_{\mathrm{hip}} - v_{\mathrm{hip}}'}f(i= \Delta v, b=[-0.5,0.5], m=1.3, b=0.1)
\end{equation}

\subsection{Standing}
We slightly modify the center of mass velocity term~$r_{v_{\mathrm{com}}^{\mathrm{xy}}}$ by setting the upper and lower bounds to 0 as $b= [0.0, 0.0]$ to penalize any movement. Besides, we add a pose tracking reward~$r_{\mathrm{pose}}$ on the local joint rotations to minimize the distance between the current pose and the designed standing pose~$\hat{q}$:
\begin{equation}
    r_{\mathrm{pose}} = exp \left[-\frac{1}{4}\sum_{j=0}^{J}||q_{j}-\hat{q}_{j}||^{2} \right],
\end{equation}
where $J$ is the number of free joints, and $q_{j}$ represents the rotation angles in radians. 

\section{Detailed Explanation of State Variables}

\begin{figure}[h!]
  \centering
  \includegraphics[width=0.4\linewidth]{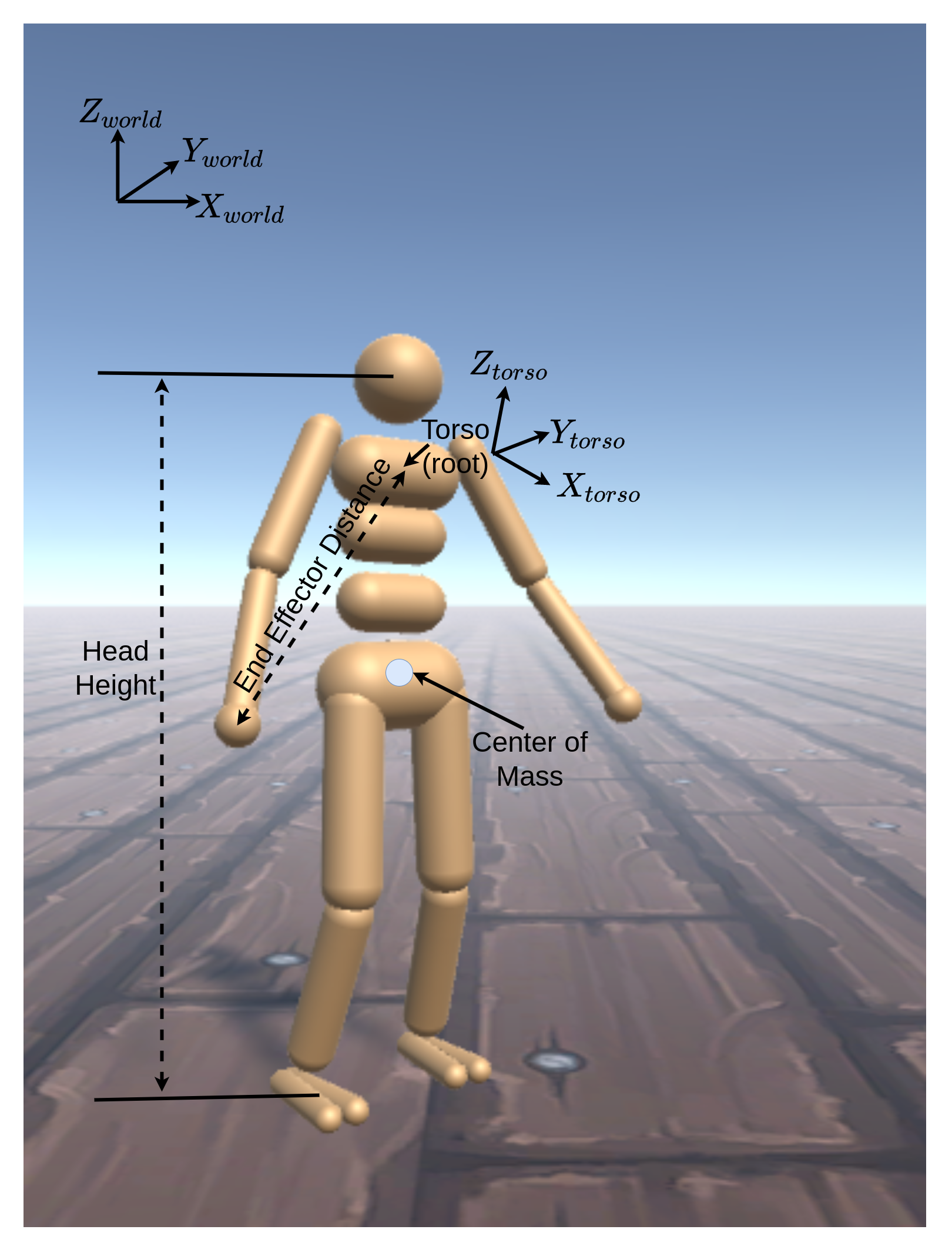}
  \caption{Diagram indicating the variables in the state definition.}
  \label{fig:state_details}
\end{figure}

As illustrated in Fig.~\ref{fig:state_details}, the head height is measured from the head link to the ground along the vertical direction. The end-effector positions are computed as the translational offset between the end-effectors and the root link in egocentric coordinates. The state also involves a variable indicating the straightness of the torso $[x_{\mathrm{torso}}^{\mathrm{up}}, y_{\mathrm{torso}}^{\mathrm{up}}, z_{\mathrm{torso}}^{\mathrm{up}}]$ as the z-component of the torso orientation vectors~$[X_{\mathrm{torso}},Y_{\mathrm{torso}},Z_{\mathrm{torso}}]$.
\section{Implementation Details}
\label{sec:sac_implementation}

All the training tasks are simulated using Mujoco engine~\cite{todorov2012mujoco} running at 800~$Hz$ while all the control policies run at 40~$Hz$. Our character is roughly 1.5 $m$ tall and weighs 38.3 $kg$. The character has a total of 19 body parts and 21 DoFs in total. The default joint angle limits and torque limits~$\mathcal{T}$ are listed in Tab.~\ref{tab:jnt_config} in App.~\ref{sec:humanoid}.

To train the controllers, we apply soft actor critic~(SAC) \cite{haarnoja2018soft} with fully connected neural networks and ReLU activation function. Controlle \piweak adopts torque as the action space while we use a low-level PD controller running at 800 $Hz$ to compute the torque for controller~$\pi_{\mathrm{slow}}$. PD controller has been proved to outperform torque as actuation model for imitation tasks~\cite{peng2017learning}. For the weak and slow get-up policies, they are trained on a 32-core CPU desktop with Nvidia Geforce RTX 2070 GPU for roughly 24 hours each. During training, policies are evaluated across ten random initial states. The ten randomly sampled initial states are fixed throughout the training.

For initial policy learning and strong-to-weak curriculum, the trained policy is evaluated every 20000 simulation steps over 10 episodes. The policy is updated through SAC every simulation step after the first 10000 simulation steps. We choose $\beta=0.95$ and $\omega=60$ throughout all the experiments. We set the clipping boundary for $\mathcal{M}_{i}$ as $N_{\mathrm{min}}=3\times10^5$ and $N_{\mathrm{max}}=8\times10^5$. For learning slower get-up motions, we choose the interpolation coefficient to be randomly sampled from $[0.2, 0.8](\kappa_{\mathrm{low}}=0.2, \kappa_{\mathrm{high}}=0.8)$ range for all the motions. In terms of the PD-controller, we set the gain to be the default joint limits $k_{p} = \mathcal{T}$, and the damping coefficient as $k_{d} = \frac{1}{10}k_{p}$. Our SAC implementation adopts the temperature annealing technique to adapt to different scales of reward functions. Both the actor and critic models are two-layer fully connected neural networks with 1024 hidden units. Tab.~\ref{tab:SAC_parameters} show the hypermeters of the SAC algorithm that we used for all the experiments.

\begin{table}[ht!]
\caption{Hyperparameters used for training the SAC algorithm.}
\label{tab:SAC_parameters}
\begin{center}
\begin{small}
\begin{tabular}{lc}
\toprule
\bf{Hyperparameter} & \bf{Value} \\
\midrule
Critic Learning Rate & $10^{-4}$ \\
Actor Learning Rate & $10^{-5}$ \\
Initial Temperature $\alpha$ & 0.1 \\
$\alpha$ Learning Rate & $10^{-4}$ \\
Optimizer & Adam \\
Log of Policy Standard Deviation Min & -5\\
Log of Policy Standard Deviation Max & 2\\
Target Update Rate ($\tau$) & $5 \cdot 10^{-3}$ \\ 
Batch Size & $1024$ \\ 
Iterations per time step & $1$ \\
\multirow{2}{*}{Discount Factor} & $0.97$ for \piweak \\
                                 & $0.95$ for \pislow \\
Reward Scaling & $1.0$ \\
Gradient Clipping & False \\
\bottomrule
\end{tabular}
\end{small}
\end{center}
\end{table}
\section{Episode Initialization Strategy Comparison}

\begin{figure}
  \centering
  \includegraphics[width=0.6\linewidth]{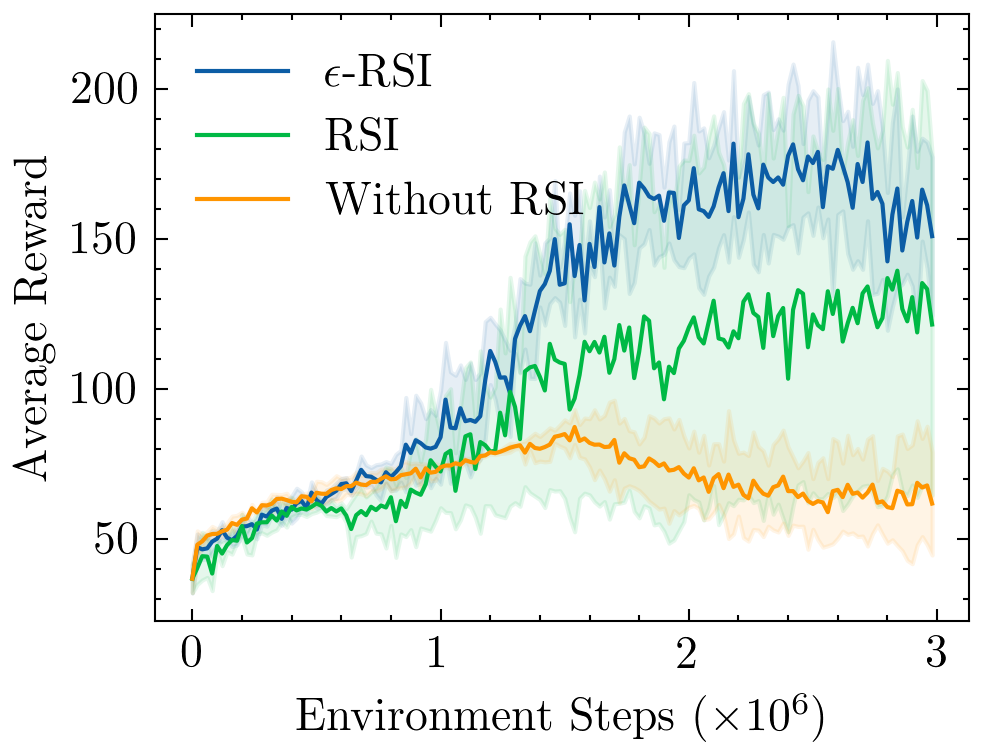}
  \caption{Training curve comparing different initialization strategies. We compare three strategies: $\epsilon$-RSI, RSI and without RSI.}
  \label{fig:rsi_comparison}
\end{figure}

In Fig.~\ref{fig:rsi_comparison}, we illustrate the performance of three initialization strategies: $\epsilon$-RSI, RSI and without RSI. Here, without RSI refers to always starting the episode from the beginning of the reference motion. Our experiment suggests that $\epsilon$-RSI outperforms other options by a large margin. Without RSI, initializing from the beginning of the reference motion can lead to instability in training. Regular RSI can successfully imitate the reference motion, but still does not perform as well as $\epsilon$-RSI.
\section{Humanoid Character Configuration}
\label{sec:humanoid}

\begin{table}
  \footnotesize
  \caption{Joint Configuration of the Humanoid Model}
  \label{tab:jnt_config}
  \begin{tabular}{lccc}
    \toprule
    Joint Name & Torque Limit~$\mathcal{T}$ (Nm) & Angle Limit (rad) & Rotation Axis\\
    \midrule
    $abdomen_{z}$     & 40  & $[-0.79, 0.79]$   & $[0,0,1]$ \\
    $abdomen_{y}$     & 40  & $[-1.31, 0.52]$   & $[0,1,0]$ \\
    $abdomen_{x}$     & 40  & $[-0.61, 0.61]$   & $[1,0,0]$ \\
    $right~hip_{x}$   & 40  & $[-0.44, 0.09]$   & $[1,0,0]$ \\
    $right~hip_{z}$   & 40  & $[-0.52, 0.44]$   & $[0,0,1]$ \\
    $right~hip_{y}$   & 120 & $[-1.92, 0.35]$   & $[0,1,0]$ \\
    $right~knee$      & 80  & $[-2.79, 0.03]$   & $[0,-1,0]$\\
    $right~ankle_{y}$ & 20  & $[-0.35, 0.79]$   & $[0,1,0]$ \\
    $right~ankle_{x}$ & 20  & $[-0.87, 0.87]$   & $[1,0,0.5]$\\
    $left~hip_{x}$    & 40  & $[-0.44, 0.09]$   & $[-1,0,0]$\\
    $left~hip_{z}$    & 40  & $[-0.52, 0.44]$   & $[0,0,-1]$\\
    $left~hip_{y}$    & 120 & $[-1.92, 0.35]$   & $[0,1,0]$\\
    $left~knee$       & 80  & $[-2.79, 0.04]$   & $[0,-1,0]$\\
    $left~ankle_{y}$  & 20  & $[-0.35, 0.79]$   & $[0,1,0]$\\
    $left~ankle_{x}$  & 20  & $[-0.87, 0.87]$   & $[1,0,0.5]$\\
    $right~shoulder1$ & 20  & $[-1.48, 1.05]$   & $[2,1,1]$\\
    $right~shoulder2$ & 20  & $[-1.48, 1.05]$   & $[0,-1,1]$\\
    $right~elbow$     & 40  & $[-1.57, 0.87]$   & $[0,-1,1]$\\
    $left~shoulder1$  & 20  & $[-1.05, 1.48]$   & $[2,-1,1]$\\
    $left~shoulder2$  & 20  & $[-1.05, 1.48]$   & $[0,1,1]$\\
    $left~elbow$      & 40  & $[-1.57, 0.87]$   & $[0,-1,-1]$\\
  \bottomrule
\end{tabular}
\end{table}

In Table~\ref{tab:jnt_config}, we list the joint torque limits and joint angle limits of the humanoid character in this work.

Our humanoid character is largely simplified if compared with the human body. Some important components are missed in the current design, including fingers and toes. Fingers and toes are vital in the get-up motion because they contact the ground directly to provide momentum to get up. In addition, the current character is modeled with simple cylinders and capsules, which is far from the shape of the real human body. We believe that careful modeling of the humanoid character can lead to more realistic motions.

\section{Details on tSNE Plots of Trajectories}
\label{sec:tsne_implementation}

Since the state variable has high dimensionality, we select the more representative features from the state variable to perform the t-SNE analysis. The selected features are ankle rotations, knee rotations, hip rotations, head height, and the vertical projection of the torso up vector $z_{\mathrm{torso}}^{\mathrm{up}}$. For each trajectory, we remove the states representing the rag-doll fall and most of the standing part. The rag-doll fall states are identical given the same initial states, which can mess up the nearest neighbour calculation for t-SNE. The same reason applies to the removal of most of the standing states. We choose the \emph{perplexity} parameter of t-SNE to be $10$ for Fig.~\ref{fig:same_controller} and Fig.~\ref{fig:different_controllers}, and $20$ for Fig.~\ref{fig:different_strategies_tsne}.
\section{Alternative Trajectory Visualization}

\begin{figure}
    \centering
    \begin{subfigure}{0.44\linewidth}
        \includegraphics[width=\linewidth]{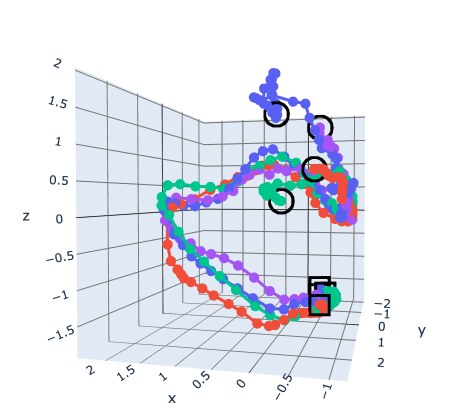}
        \caption{Different Initial States}
        \label{fig:same_controller_pca}
    \end{subfigure}
    \begin{subfigure}{0.49\linewidth}
        \includegraphics[width=\linewidth]{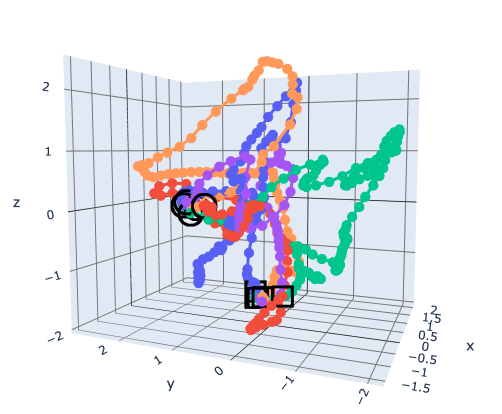}
        \caption{Different Controllers}
        \label{fig:different_controllers_pca}
    \end{subfigure}
    
    \caption{PCA plots of trajectories. These two diagrams correspond to the same data but different dimensionality reduction algorithm of Fig.~\ref{fig:same_controller} and Fig.~\ref{fig:different_controllers}.}
    \label{fig:pca_plots}
\end{figure}

We choose t-SNE algorithm to compress the high dimension state to a 3D vector for visualization purpose. Alternatively, Principal Component Analysis~(PCA) serves as another popular dimensionality reduction algorithm to encode the original state variables through a linear transformation. As shown in Fig.~\ref{fig:pca_plots}, we experiment the choice of PCA on the same data used in the visualization demonstrated in Fig.~\ref{fig:tsne_plots}. The plots generated by PCA still provide reasonable trajectories. However, the converging behavior of the same controller is not as clearly illustrated as with t-SNE. Additionally, different trajectories produced by different controllers spread out, and are more difficult to interpret.
\end{document}